# 2D condensate of electrons and holes in ultrathin MoTe$_2$ photocells

Trevor B. Arp[1,2], Dennis Pleskot[2,3], Vivek Aji[1], Nathaniel M. Gabor[1,2,3,4*]

[1]Department of Physics and Astronomy, University of California, 900 University Avenue, Riverside, California 92521, United States

[2]Laboratory of Quantum Materials Optoelectronics, Materials Science and Engineering Building, Room 179, University of California, Riverside, California 92521, United States

[3]Department of Materials Science and Engineering, University of California, 900 University Avenue, Riverside, California 92521, United States

[4]Canadian Institute for Advanced Research, CIFAR Azrieli Global Scholar, MaRS Centre West Tower, 661 University Avenue, Toronto, Ontario ON M5G 1M1, Canada

*Correspondence to: nathaniel.gabor@ucr.edu

**Abstract:** The electron-hole liquid, which features a macroscopic population of correlated electrons and holes, may offer a path to room temperature semiconductor devices that harness collective electronic phenomena. We report on the gas-to-liquid phase transition of electrons and holes in ultrathin molybdenum ditelluride photocells revealed through multi-parameter dynamic photoresponse microscopy (MPDPM). By combining rich visualization with comprehensive analysis of very large data sets acquired through MPDPM, we find that ultrafast laser excitation at a graphene-MoTe$_2$-graphene interface leads to the abrupt formation of ring-like spatial patterns in the photocurrent response as a function of increasing optical power at $T$ = 297 K. These patterns, together with extreme sublinear power dependence and picosecond-scale photocurrent dynamics, provide strong evidence for the formation of a two-dimensional electron-hole condensate.

Condensation - the familiar process underlying the formation of clouds and the distillation of ethyl alcohol into whiskey - is the phase transition of gas into liquid (*1*). In semiconductors, non-equilibrium charge carriers exist as a gas of free electrons and holes, bound electron-hole pairs (excitons), biexcitons, and trions (charged excitons) (*2-5*). Remarkably, at sufficiently high electron-hole (*e-h*) densities or low temperatures, the non-equilibrium *e-h* system may undergo condensation (*6-10*). Negatively charged electrons (*e⁻*) and positively charged holes (*h⁺*) merge to become an electronic liquid.

Ultrathin $MoTe_2$ is an ideal semiconductor for the study of collective electron-hole phases, yet no measurements on two-dimensional (2D) transition metal dichalcogenides (TMDs) have clearly demonstrated the gas-to-liquid phase transition. Very high *e-h* densities have been demonstrated in TMDs using ultrashort laser pulses, giving rise to giant bandgap renormalization (*11,12*) and strong exciton-exciton interactions (*13,14*). By integrating multi-layer $MoTe_2$ into graphene-$MoTe_2$-graphene vertical heterostructures, such strong many-body effects may be accessed in the ultrasensitive optoelectronic response (*15-18*), which results from the indirect band gap and long exciton lifetimes (*19-21*). As photocell device complexity increases, however, efficient photoresponse techniques must be developed to observe, manipulate, and harness novel 2D electronic phases.

Here, we report on a data-intensive optoelectronic imaging technique - called multi-parameter dynamic photoresponse microscopy (MPDPM) - that reveals a room temperature 2D electron-hole condensate in ultrathin $MoTe_2$ photocells. By incorporating ultrafast optoelectronic measurements (*22-27*) with efficient data acquisition, automation and analysis, we rapidly and densely sample a broad experimental parameter space and explore data correlations at various length and time scales (details in supplemental materials section S1). MPDPM utilizes a near-infrared ultrafast laser, in which ultrashort (150 fs) laser pulses are split into two distinct paths. A translation stage controls the path length difference between the split pulses. The two identical-power beams are then recombined and the path length difference of the delayed beam results in a time delay between pulses $\Delta t$ (Figure 1A). The recombined beam is focused to a diffraction-limited spot that is spatially scanned across the 2D optoelectronic devices under high vacuum (supplemental materials section S1.1).

Using dynamic photoresponse microscopy, we acquire a multidimensional data set of interlayer photocurrent $I$ vs. two spatial dimensions, laser power $P$, time delay $\Delta t$, and interlayer voltage $V_i$ (supplemental materials section S1.3). Photocurrent generated across a graphene-$MoTe_2$-graphene photocell (Figure 1A) is measured at each point in space to generate a map of the interlayer photocurrent response (Figure 1B). This imaging process is repeated as a function of increasing optical power (supplementary movie 1), while incrementing $\Delta t$ and interlayer voltage $V_i$. Figure 1C shows that the interlayer photocurrent increases sub-linearly with increasing power at $\Delta t = 50$ ps. Such photoresponse is consistent with previous photocurrent measurements of TMD photocells (*15,16,26,27*): photogenerated electron-hole pairs are promoted across the indirect band gap of $MoTe_2$ and collected as individual electrons and holes (Figure 1C inset).

Our data-intensive technique allows us to gain a comprehensive understanding of the ordinary $MoTe_2$ heterostructure photoresponse. The photocurrent power dependence (Figure 1C) is well described by a single power law, $I \propto P^\gamma$, where the power law exponent $\gamma = 0.44$ parameterizes the nonlinearity of the photoresponse. A power law exponent $\gamma \sim 1/2$ suggests straightforward dynamics with a simple rate equation $dN/dt = -N/\tau_{esc} - \alpha N^2$, where $N$ is the *e-h* pair density, $\tau_{esc}$ is the carrier escape time, and $\alpha$ is the exciton-exciton annihilation rate (*25*) (supplemental materials section S4.3). By including a constant generation rate, the steady state ($dN/dt = 0$) solution to this rate equation results in photocurrent $I \propto P^{1/2}$, in good agreement with the observed $I \propto P^{0.44}$. Since MPDPM uses ultrashort pulses, however, careful time integration of the dynamics is required, resulting in an analytic solution $I \propto \ln(1 + N_0 \alpha \tau_{esc})/\alpha \tau_{esc}$, where $N_0$ is the *e-h* density immediately following the laser pulse (supplemental materials section S4.3). The analytic solution (red dashed line Figure 1C) exhibits excellent agreement with the photocurrent data (blue data) and the power law fit (black line). We conclude that $\gamma$ is thus a robust parameterization of the nonlinear photoresponse.

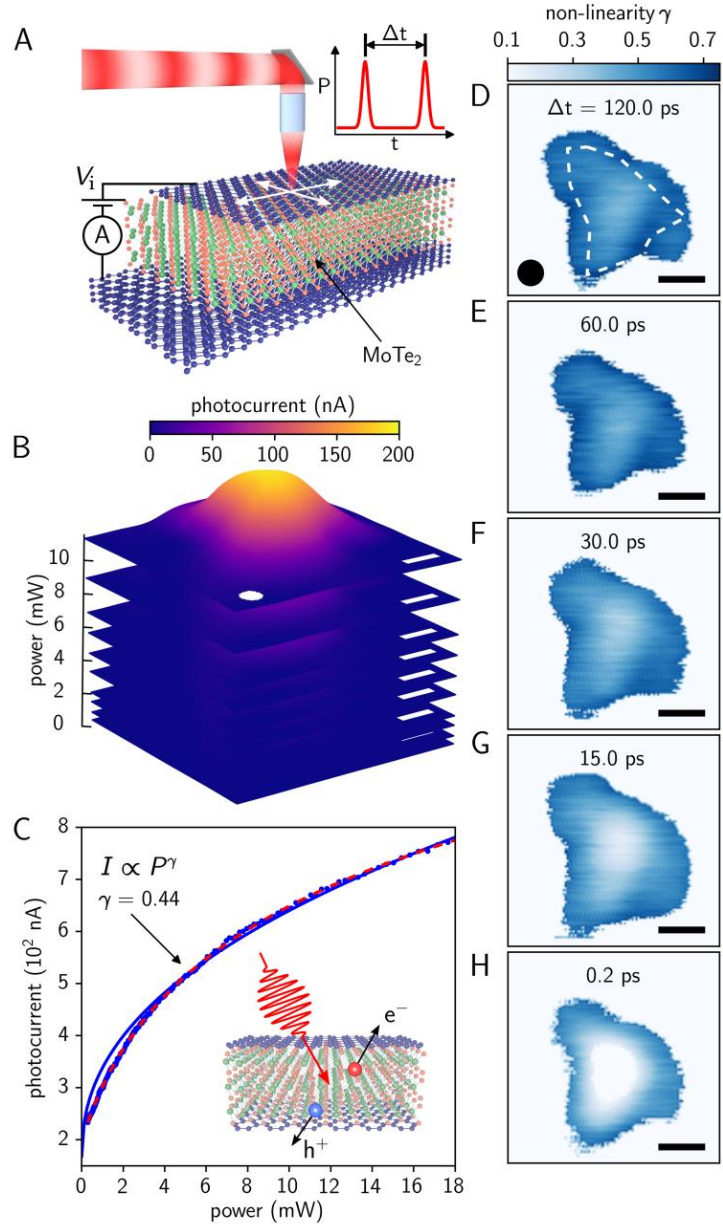

**Fig. 1**. *Multi-parameter dynamic photoresponse microscopy (MPDPM) of ultrathin MoTe₂ photocells.* (**A**). Schematic of the photocell and measurement. (**B**) MPDPM images for increasing laser power; wavelength $\lambda$ = 1200 nm, $T$ = 297 K, and time delay $\Delta t$ = 120 ps. (**C**) Photocurrent vs. optical power in the center of the heterostructure (blue data) of thickness 9 nm. Solid blue line is a fit to a power law, and red dashed line is the analytical model fit. Inset, schematic of photoexcitation and collection of electrons (e⁻) and holes (h⁺) in MoTe₂. (**D-H**) MPDPM images of the power law exponent, $\gamma(x, y)$, as a function of two-pulse time delay $\Delta t$ (labeled). Dashed line in D outlines the graphene-MoTe₂-graphene heterostructure. Scale bars 3 $\mu$m. Circles indicate the full width at half maximum of the diffraction-limited beamspot.

The multi-parameter dynamic photoresponse microscope visualizes the nonlinear photoresponse with extraordinary spatio-temporal detail (Figures 1D-H). From a large set of photocurrent images (as in Figure 1B), the interlayer photocurrent vs. optical power is fit to $I \propto P^{\gamma}$ at each point in space. This large data set is condensed into an image of the photocurrent nonlinearity $\gamma(x,y)$, which we then measure as a function of $\Delta t$ (supplementary movie 2). A snapshot from the spatio-temporal dynamics at $\Delta t = 120$ ps (Figure 1D) shows that the nonlinearity $\gamma(x,y)$ is nearly uniform over the active area of the MoTe$_2$ heterostructure (dashed outline Figure 1D), exhibiting a narrow range $\gamma = 0.45$ - 0.60. The spatially uniform photoresponse with $I \propto P^{1/2}$ at long time delay is fully consistent with ordinary photoresponse due to exciton-exciton interactions (*13,14,25*).

Strikingly, when the time delay between laser pulses is very short, MPDPM reveals highly anomalous photoresponse. At $\Delta t = 0.2$ ps (Figure 1H), the power law behavior collapses near the center, resulting in a pronounced ring of sublinear photoresponse ($\gamma \sim 0.5$). Figures 1F,G show that the sudden collapse near $\Delta t = 0.2$ ps is preceded by a gradual suppression of $\gamma$ at longer time delays. The area of power law suppression significantly exceeds the beam spot size, indicating a global change in photoresponse. In the following, we examine the space-time evolution of the MoTe$_2$ photoresponse, and extract detailed dependence of the spatial photocurrent features on optical power, interlayer voltage, and time delay.

We first decompose the MPDPM image measured at $\Delta t = 0.2$ ps (Figure 1H) and examine the constituent photocurrent maps (Figure 2). At low optical powers, the photocurrent magnitude increases rapidly and monotonically (Figure 2A). For $P > 6$ mW, however, the photocurrent at the center of the device suddenly decreases, forming a photocurrent ring of bright photoresponse (see supplementary movie 3). The photocurrent ring grows rapidly with increasing optical power. To see the ring expansion more clearly, Figure 2B shows the magnitude of the spatial gradient of the photocurrent maps $|\nabla I|$, which we use to visualize the local slope of the spatially resolved photocurrent landscape. At an optical power $P = 6$ mW, a clear edge begins to emerge and grows into a well-formed ring.

Remarkably, the anomalous photocurrent ring appears abruptly with increasing optical power. Using the gradient maps (Figure 2B), we quantify the ring area by algorithmically identifying the contour $|\nabla I| \approx 0$ (see supplemental materials section S1.5 for details). Figure 2C shows the power dependence of the ring volume (product of area and sample thickness) as a fraction of the total heterostructure volume. At a critical power $P_C = 6$ mW, we observed a nearly discontinuous growth rate of the volume fraction. Above the transition $P > P_C$, the photocurrent ring, and thus volume fraction, expands linearly with optical power.

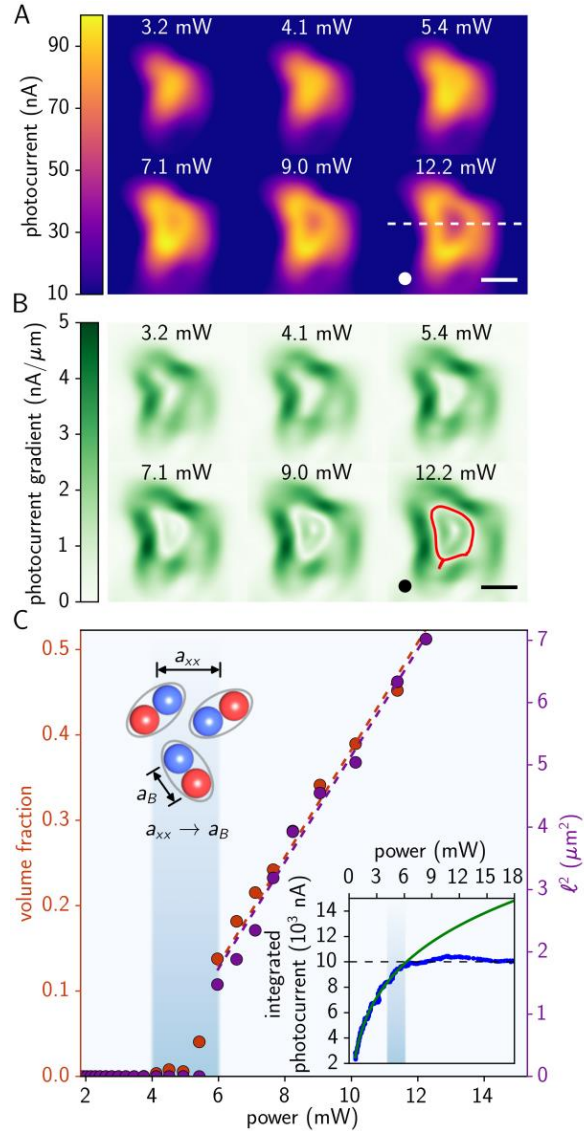

**Fig. 2.** *Critical onset of ring-like photoresponse revealed through MPDPM.* (**A**) Spatially resolved photocurrent measured at various powers (labeled) and $\Delta t = 0.2$ ps. Dashed line indicates location of photocurrent line profiles. (**B**) Photocurrent gradient $|\nabla I|$ calculated from photocurrent maps in A. The contour $|\nabla I| \approx 0$ encloses the photocurrent ring (red contour in the image at $P = 12.2$ mW). Scale bars 5 $\mu$m, circles indicate the beamspot FWHM. (**C**) Photocurrent ring volume fraction (red data) vs. laser power. Volume fraction is the ratio of the volume enclosed by the $|\nabla I| \approx 0$ contours to the active photocell volume (area). Peak-to-valley distance $\ell$ (purple data) vs. optical power extracted from photocurrent line cuts in A. Red (purple) dashed lines are linear fits to volume fraction ($l^2$) vs. optical power above $P = 6$ mW. Inset top, inter-exciton spacing $a_{xx}$ approaches the exciton Bohr radius $a_B$. Inset bottom, spatially integrated photocurrent vs. power (blue data) and power law fit below 6 mW (solid green line).

The sharp transition at $P_C$ also manifests as a sudden deviation from power law behavior. Figure 2C lower right inset shows the spatially integrated photocurrent vs. power measured along the dashed line in Figure 2A. The photocurrent increases rapidly at low power, and exhibits ordinary power law growth (solid green line Figure 2C lower right inset). Above $P = P_C$, however, the data falls significantly below the power law fit, and the spatially integrated photocurrent remains nearly constant as power increases. Thus, the abrupt formation and expansion of the photocurrent ring corresponds directly to the collapse of power law behavior observed in Figure 1H.

We attribute the anomalous photoresponse in graphene-MoTe$_2$-graphene photocells to spontaneous condensation of a 2D electron-hole liquid. At low laser power, photo-excitation generates a gas of electrons and holes in MoTe$_2$ (Figure 3A). Enhanced Coulomb interactions bind electrons and holes into excitons with nanometer-scale Bohr radius $a_B$ (6-8). Due to long $e$-$h$ pair lifetimes, the $e$-$h$ pair density confined within MoTe$_2$ increases with laser power until exciton-exciton interactions become comparable to interactions within an individual bound $e$-$h$ pair. Below the power threshold $P_C$, ordinary two-body ($N^2$) recombination processes dominate the interlayer photocurrent $I \propto P^{1/2}$ (Figure 1C).

At the critical laser power $P_C$, the electron-hole population merges into a many-body condensate (Figure 3A). The $e$-$h$ pair density $N$ becomes so large that the average spacing between pairs is nearly equal to the exciton radius (Figure 2C upper left inset). At $P_C = 6$ mW, the mean exciton-exciton separation, which we estimate to be $a_{xx} = 1$-3 nm in MoTe$_2$ (supplemental materials section S4.1), is very close to the Bohr radius $a_B = 2.3$ nm extracted from magneto-optical measurements (28). Once $a_{xx} \sim a_B$, the $e$-$h$ population reaches the critical density $N_C \sim 0.5/nm^3$ $e$-$h$ pairs, and many-body effects dominate. Above the phase transition, the renormalization of the energy per $e$-$h$ pair results in a suppression of photon absorption within the $e$-$h$ liquid (11,12). The resulting 2D condensate exhibits a fixed electron-hole pair density $N_C$, is highly polarizable in an applied electric field, and forms a sharp, stable boundary that separates it from the gas phase (6-8).

The 2D electron-hole condensate exhibits several important features, which are readily revealed through MPDPM. First, $e$-$h$ condensation results in highly unusual ring-like interlayer photoresponse, which arises from the convolution of the imaging beam spot with a sharply bound area of suppressed absorption (supplemental materials section S4.2). Figure 3B and 3C compare the $e$-$h$ liquid model to interlayer photocurrent line traces as a function of increasing power (measured along the dashed line in Figure 2A). As expected, the squared center-to-edge distance of the ring $l^2$ increases linearly above the critical threshold (purple data Figure 2C), exhibiting nearly identical growth to the volume fraction. The $e$-$h$ liquid model (Figure 3B) shows excellent agreement with the photocurrent line profiles: Above the phase transition, energy added to the condensate contributes exclusively to expansion of the $e$-$h$ liquid in MoTe$_2$.

The bound electron-hole liquid is strongly polarizable in an applied electric field. For interlayer voltages $V_i$ above the built-in potential $\phi_0 = -41$ mV (supplemental materials section S3.1), the center-to-edge distance $l^2$ decreases approximately linearly as voltage increases (Figure 4A). When the total interlayer voltage exceeds the critical voltage $eV_C = e(V_i - \phi_0) > 45$ meV, electrons and holes evaporate from the *e-h* liquid and become ordinary *e-h* pairs. Above the critical interlayer voltage $V_C$, spatially uniform photocurrent re-emerges as the 2D condensate dissociates in the electric field (supplemental materials section 3.2).

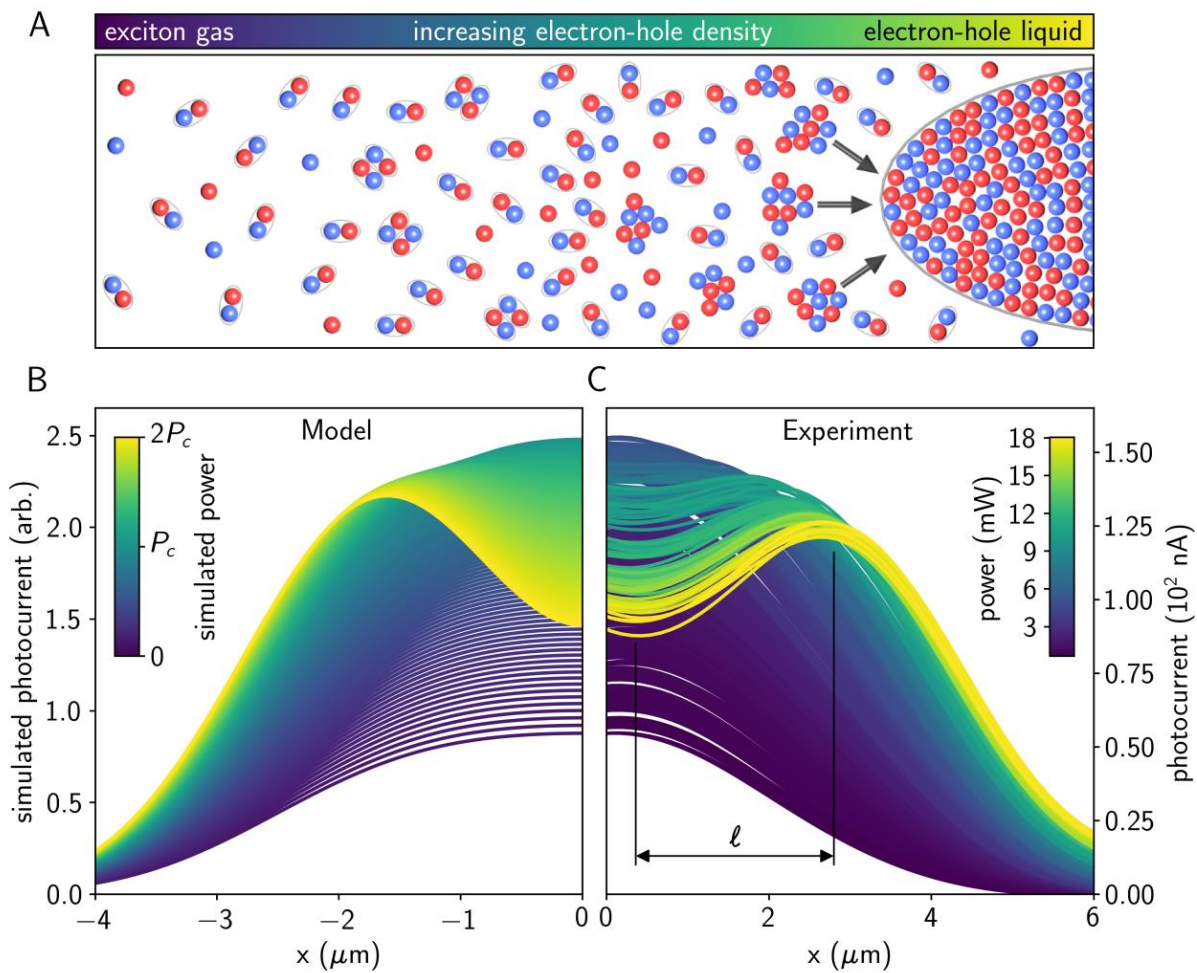

**Fig. 3.** *Room temperature 2D electron-hole condensate and comparison to MPDPM imaging in the MoTe₂ photocells.* (**A**) Evolution of electron-hole interactions with increasing *e-h* density. As density increases, the non-interacting gas of excitons gives rise to exciton-exciton interactions, eventually leading to condensation into a 2D electron-hole liquid. (**B**) Calculated spatially resolved photoresponse of the electron-hole condensate showing the suppression of photocurrent in the center of the sample above the critical power $P_c$. (**C**) Photocurrent line profiles measured across the center of the sample for increasing power; $T$ = 297 K, $\Delta t$ = 0.2 ps.

MPDPM reveals the dynamic transition between the electron-hole liquid and gas phase. Figures 4B, 4C show the time-resolved photocurrent and photocurrent nonlinearity $\gamma$. When the laser is fixed at the center of the device, the photocurrent vs. $\Delta t$ exhibits remarkably different power dependence between short and long time delay. At short $\Delta t$, the photocurrent at the center of the device decreases with increasing power above $P_C$ (supplemental materials section S3.2). This extreme sublinear photoresponse is fully consistent with power law collapse associated with condensation (Fig. 2C lower right inset). At long time delay, the photocurrent exhibits ordinary gas-phase behavior $I \propto P^{0.52}$ (Figure 4C). We fit $\gamma$ vs. $\Delta t$ to an exponential decay (black line Figure 4C) to extract the charge density persistence time $\tau = 22$ ps. For $\Delta t > \tau$ the pulses are independent and each is insufficient to drive condensation. Thus, photoexcitation produces a gas of ordinary e-h pairs. For $\Delta t < \tau$ the combined charge density produced by the two pulses is sufficient to cause the gas-to-liquid transition. Numerically modeling the detailed dynamics of free charge carriers, excitons, and *e-h* pairs in the device reproduces this behavior (see supplemental materials section S4.3).

Electron-hole condensation at room temperature is a surprising result, and may make possible TMD optoelectronic devices that harness electronic fluids under ordinary operating conditions. The gas-to-liquid phase transition is set by the energy difference $\Delta E$ between the average energy per *e-h* pair in the gas phase and the reduced energy per *e-h* pair in the liquid phase (supplemental materials section S4.3). When $\Delta E$ is large compared to thermal energy at room temperature ($K_B T_{300K} = 26$ meV), the condensate is stable against thermal fluctuations. From the interlayer voltage dependence (Figure 4A), we estimate that $\Delta E \sim eV_C \sim 45$ meV, approximately twice the thermal energy at room temperature. While this renormalization is comparable to conventional 2D electron systems, the *e-h* pair binding energy ($\sim 10^2$ meV) in TMDs is several orders of magnitude larger (*11,12,29,30*). In 2D TMDs, the large binding energy and strong exciton-exciton interactions combine to allow for condensation at room temperature.

**Acknowledgments:** The authors would like to acknowledge valuable discussions with Chandra Varma, James Hone, Frank Koppens, and Mathieu Massicotte. This work was supported by the Air Force Office of Scientific Research Young Investigator Program (YIP) award # FA9550-16-1-0216, as part of the SHINES center, an Energy Frontier Research Center funded by the U.S. Department of Energy, Office of Science, Basic Energy Sciences under award no. SC0012670, and through support from the National Science Foundation Division of Materials Research CAREER award no. 1651247. D.P. and N.M.G received support from SHINES. N.M.G. acknowledges support through a Cottrell Scholar Award, and through the Canadian Institute for Advanced Research (CIFAR) Azrieli Global Scholar Award. T.B.A. acknowledges support from the Fellowships and Internships in Extremely Large Data Sets (FIELDS) program, a NASA MUREP Institutional Research Opportunity (MIRO) program, grant number NNX15AP99A.

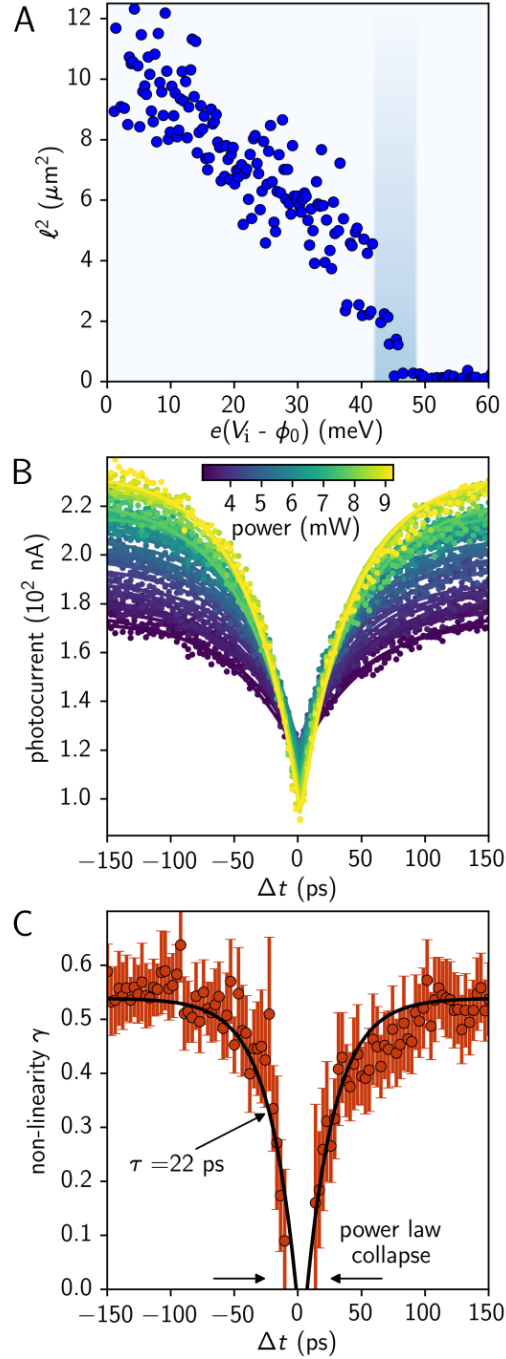

**Fig. 4.** *Interlayer voltage dependence and dynamic photoresponse of the 2D electron-hole condensate.* (**A**) Photocurrent ring peak-to-valley distance $l^2$ vs. interlayer voltage; $\Delta t = 0.2$ ps. $\phi_0$ is the built-in potential of the graphene-MoTe$_2$-graphene photocell. (**B**) Photocurrent vs. $\Delta t$ for increasing optical power. The laser is fixed at the center of the device. Solid lines are exponential fits to the data at each power. (**C**) Power law exponent $\gamma$ as a function of $\Delta t$, extracted by fitting the data in B to $I \propto P^\gamma$. Solid black line is an exponential fit with a characteristic time-scale $\tau = 22$ ps.

**References:**

1. R. Feynman, R. Leighton, M. Sands, *The Feynman Lectures on Physics*, (Addison-Wesley, Reading, MA, 1963), pp. 40-3.

2. M. Lampert, Mobile and immobile effective-mass-particle complexes in nonmetallic solids. *Phys. Rev. Lett.* **1**, 450–453 (1958).

3. K. Kheng, R. Cox, M. D'Aubigné, Observation of negatively charged excitons X− in semiconductor quantum wells. *Phys. Rev. Lett.* **71**, 1752–1755 (1993).

4. G. D. Scholes, G. Rumbles, Excitons in nanoscale systems. *Nat. Mater.* **5**, 683–696 (2006).

5. K. F. Mak, K. He, C. Lee, G. H. Lee, J. Hone, T. F. Heinz, J. Shan, Tightly bound trions in monolayer $MoS_2$. *Nature Materials* **12**, 207–211 (2013).

6. L. V. Keldysh, in *Proceedings of the $9^{th}$ International Conference on Physics of Semiconductors*, (Nauka Leningrad, 1968) 1303.

7. C. D. Jeffries, Electron-hole condensation in semiconductors. *Science* **189**, 955-964 (1975).

8. L. V. Keldysh, The electron-hole liquid in semiconductors. *Contemp. Phys.* **27**, 395-428 (1986).

9. R. A. Kaindl, M. A. Carnahan, D. Hägele, R. Lövenich, D. S. Chemla, Ultrafast terahertz probes of transient conducting and insulating phases in an electron–hole gas. *Nature* **423**, 734-738 (2003).

10. A. E. Almand-Hunter, H. Li, S. T. Cundiff, M. Mootz, M. Kira, S. W. Koch, Quantum droplets of electrons and holes. *Nature* **506**, 471-475 (2014).

11. M. M. Ugeda, A. J. Bradley, S. Shi, F. H. da Jornada, Y. Zhang, D. Y. Qiu, W. Ruan, S. Mo, Z. Hussain, Z. Shen, F. Wang, S. G. Louie, M. F. Crommie, Giant bandgap renormalization and excitonic effects in a monolayer transition metal dichalcogenide semiconductor. *Nature Materials* **13**, 1091-1095 (2014).

12. A. Chernikov, C. Ruppert, H. M. Hill, A. F. Rigosi, T. F. Heinz, Population inversion and giant bandgap renormalization in atomically thin $WS_2$ layers. *Nature Photonics* **9**, 466-470 (2015).

13. D. Sun, Y. Rao, G. A. Reider, G. Chen, Y. You, L. Brézin, A. R. Harutyunyan, T. F. Heinz, Observation of Rapid Exciton–Exciton Annihilation in Monolayer Molybdenum Disulfide. *Nano Lett.*, **14**, 5625-5629 (2014).

14. G. Froehlicher, E. Lorchat, S. Berciaud, Direct versus indirect band gap emission and exciton-exciton annihilation in atomically thin molybdenum ditelluride $MoTe_2$. *Phys. Rev. B* **94**, 085429 (2016).

15. K. Zhang, X. Fang, Y. Wang, Y. Wan, Q. Song, W. Zhai, Y. Li, G. Ran, Y. Ye, L. Dai, Ultrasensitive near-infrared photodetectors based on a graphene-$MoTe_2$-graphene vertical Van der Waals heterostructure. *ACS Applied Materials and Interfaces* **9**, 5392-5398 (2017).

16. F. Wang, L. Yin, Z. Wang, K. Xu, F. Wang, T. A. Shifa, Y. Huang, Y. Wen, C. Jiang, J. He, Strong electrically tunable $MoTe_2$-graphene Van der Waals heterostructures for high-performance electronic and optoelectronic devices. *Applied Physics Letters* **109**, 193111 (2016).

17. M. Kuiri, B. Chakraborty, A. Paul, S. Das, A. K. Sood, A. Das, Enhancing photoresponsivity using $MoTe_2$-graphene vertical heterostructures. *Applied Physics Letters* **108**, 063506 (2016).

18. T. J. Octon, V. K. Nagareddy, S. Russo, M. F. Craciun, C. D. Wright, Fast High-Responsivity Few-Layer MoTe$_2$ Photodetectors. *Advanced Optical Materials,* **4**, 1750-1754 (2016).

19. C. Ruppert, O. B. Aslan, T. F. Heinz, Optical properties and band gap of single and few-layer MoTe$_2$ crystals. *Nano Letters* **14**, 6231-6236 (2014).

20. I. G. Lezama, A. Arora, A. Ubaldini, C. Barreteau, E. Giannini, M. Potemski, A. F. Morpurgo, Indirect-to-direct band gap crossover in few-layer MoTe$_2$. *Nano Letters* **15**, 2336-2342 (2015).

21. G. P. Kekelidzet, B. L. Evans, The photovoltage in single crystals of a-MoTe$_2$. *Brit. J. Appl. Phys.* **2**, 855-861 (1969).

22. N. M. Gabor, Z. Zhong, K. Bosnick, P. L. McEuen, Ultrafast Photocurrent Measurement of the Escape Time of Electrons and Holes from Carbon Nanotube p−i−n Photodiodes. *Phys. Rev. Lett.* **108**, 087404 (2012).

23. H. Wang, C. Zhang, W. Chan, S. Tiwari, F. Rana, Ultrafast response of monolayer molybdenum disulfide photodetectors. *Nat. Comm.* **6**, 8831 (2015).

24. Q. Ma, T. I. Andersen, N. L. Nair, N. M. Gabor, M. Massicotte, C. H. Lui, A. F. Young, W. Fang, K. Watanabe, T. Taniguchi, J. Kong, N. Gedik, F. H. L. Koppens, P. Jarillo-Herrero, Tuning ultrafast electron thermalization pathways in a van der Waals heterostructure. *Nat. Phys.* **12** 455-459 (2016).

25. K. T. Vogt, S. Shi, F. Wang, M. Graham, Isolating Exciton Extraction Pathways with Electric Field-Dependent Ultrafast Photocurrent Microscopy. *Conference on Lasers and Electro-Optics,* (Optical Society of America Technical Digest, 2016).

26. M. Massicotte, P. Schmidt, F. Vialla, K. G. Schdler, A. Reserbat-Plantey, K. Watanabe, T. Taniguchi, K. J. Tielrooij, F. H. L. Koppens, Picosecond photoresponse in Van der Waals heterostructures. *Nature Nanotechnology* **11**, 42-46 (2016).

27. M. Massicotte, P. Schmidt, F. Vialla, K. Watanabe, T. Taniguchi, K. J. Tielrooij, F. H. L. Koppens, Photo-thermionic effect in vertical graphene heterostructures. *Nat. Comm.* **7**, 12174 (2016).

28. Y. Sun, J. Zhang, Z. Ma, C. Chen, J. Han, F. Chen, X. Luo, Y. Sun, Z. Sheng, The Zeeman splitting of bulk 2H-MoTe$_2$ single crystal in high magnetic field. *Appl. Phys. Lett.* **110**, 102102 (2017).

29. A. Chernikov, T. C. Berkelbach, H. M. Hill, A. Rigosi, Y. Li, O. B. Aslan, D. R. Reichman, M. S. Hybertsen, T. F. Heinz, Exciton binding energy and nonhydrogenic rydberg series in monolayer WS$_2$. *Phys. Rev. Lett.* **113**, 076802 (2014).

30. A. Arora, R. Schmidt, R. Schneider, M. R. Molas, I. Breslavetz, M. Potemski, R. Bratschitsch, Valley Zeeman Splitting and Valley Polarization of Neutral and Charged Excitons in Monolayer MoTe$_2$ at High Magnetic Fields. *Nano Lett.* **16** 3624–3629 (2016).

31. G. Steinmeyer, A review of ultrafast optics and optoelectronics. *Journal of Optics A: Pure and Applied Optics* **5**, R1-R15 (2003).

32. M. Fushitani, Applications of pump-probe spectroscopy. *Annu. Rep. Prog. Chem., Sect. C: Phys. Chem.* **104**, 272 (2008).

33. E. Hecht, *Optics*. (Addison-Wesley, Reading, MA, 2002).

34. W. F. Brinkman, T. M. Rice, P. W. Anderson, S. T. Chui, Metallic state of the electron-hole liquid, particularly in germanium. *Phys. Rev. Lett.* **28**, 961-964 (1972).

35. L. V. Keldysh, Coulomb interaction in thin semiconductor and semimetal films. *Soviet Journal of Experimental and Theoretical Physics Letters* **29**, 658 (1979).

36. I. G. Lezama, A. Ubaldini, M. Longobardi, E. Giannini, C. Renner, A. B. Kuzmenko, A. F. Morpurgo, Surface transport and band gap structure of exfoliated 2H-MoTe₂ crystals. *2D Materials* **1**, 021002 (2014).

37. A. Castellanos-Gomez, M. Buscema, R. Molenaar, V. Singh, L. Janssen, H. S J van der Zant, G. A. Steele, Deterministic transfer of two-dimensional materials by all-dry viscoelastic stamping. *2D Mater. Lett.* **1**, 1-8 (2014).

38. A. C. Ferrari, J. C. Meyer, V. Scardaci, C. Casiraghi, M. Lazzeri, F. Mauri, S. Piscanec, D. Jiang, K. S. Novoselov, S. Roth, A. K. Geim, Raman spectrum of graphene and graphene layers. *Phy. Rev. Lett.,* **97**, 187401 (2006).

39. O. P. Agnihotri, H. K. Sehgal, A. K. Garg, Laser Excited Raman Spectra of Gr. VI Semiconducting Compounds. *Solid State Communications,* **12**, 135-138 (1973).

40. M. Yamamoto, S. T. Wang, M. Ni, Y. Lin, S. Li, S. Aikawa, W. Jian, K. Ueno, K. Wakabayashi, K. Tsukagoshi, Strong enhancement of Raman scattering from a bulk-inactive vibrational mode in few-layer MoTe₂. *ACS Nano,* **8**, 3895-3903 (2014).

41. G. Cunningham, M. Lotya, C. S. Cucinotta, S. Sanvito, S. D. Bergin, R. Menzel, M. S. P. Shaffer, J. N. Coleman, Solvent Exfoliation of Transition Metal Dichalcogenides: Dispersibility of Exfoliated Nanosheets Varies Only Weakly between Compounds. *ACS Nano,* **6**, 3468-3480 (2012).

42. S. Cho, S. Kim, J. H. Kim, J. Zhao, J. Seok, D. H. Keum, J. Baik, D. Choe, K. J. Chang, K. Suenaga, S. W. Kim, Y. H. Lee, H. Yang, Phase patterning for ohmic homojunction contact in MoTe₂. *Science,* **349**, 625-628 (2015).

43. N. R. Pradhan, D. Rhodes, S. Feng, Y. Xin, S. Memaran, B. Moon, H. Terrones, M. Terrones, L. Balicas, Field-Effect Transistors Based on Few-Layered R-MoTe₂. *ACS Nano,* **8**, 5911-5920 (2014).

44. W. J. Yu, Y. Liu, H. Zhou, A. Yin, Z. Li, Y. Huang, X. Duan, Highly efficient gate-tunable photocurrent generation in vertical heterostructures of layered materials. *Nature Nanotechnology,* **8**, 952-958 (2013).

45. Y. Lin, Y. Xu, S. Wang, S. Li, M. Yamamoto, A. Aparecido-Ferreira, W. Li, H. Sun, S. Nakaharai, W. Jian, K. Ueno, K. Tsukagoshi, Ambipolar MoTe₂ transistors and their applications in logic circuits. *Advanced Materials* **26**, 3263-3269 (2014).

46. R. Boker, Band Structure of MoS₂ , MoSe₂, and α-MoTe₂: Angle-Resolved Photoelectron Spectroscopy and Ab Initio Calculations. *Physical Review B,* **64**, 235305 (2001).

# Supplementary Materials for
# 2D condensate of electrons and holes in ultrathin MoTe$_2$ photocells

The molybdenum ditelluride heterostructures studied in this work were assembled using exfoliated bulk transition metal dichalcogenide (TMD) crystals and novel annealing and fabrication processes to explore intrinsic photoresponse and the 2D electron-hole condensate. The following supplement contains detailed characterization of two devices; each was assembled and annealed using novel techniques, described herein. Extensive optical spectroscopic (including Raman, photoluminescence, differential reflection, and photocurrent spectroscopy) and optoelectronic characterization confirmed strong photoresponse and ordinary multi-exciton dynamics in the MoTe$_2$ photocells.

In Section S1 we discuss our data-intensive optoelectronic imaging technique – multiple parameter dynamic photoresponse microscopy (MPDPM) - used to generate the highly detailed photoresponse images shown in the main text. In Section S2 we describe the fabrication and characterization of the graphene-MoTe$_2$-graphene photocells studied using MPDPM. Section S3 describes the optoelectronic characterization of our samples, which is used to demonstrate the highly temperature-sensitive and wavelength-dependent photoresponse of MoTe$_2$ photocells. This establishes baseline optoelectronic behavior. In addition, Section S3 provides a detailed description of the photoresponse data at the threshold of *e-h* condensation, expanding on data which is simplified in the main text figures. Section S4 details calculations of the 2D *e-h* condensate model which are referenced throughout the main text.

## S1. Multi-Parameter Dynamic Photoresponse Microscopy

Multi-Parameter Dynamic Photoresponse Microscopy (MPDPM) is a new technique we have developed for probing the dynamics of charge carriers in heterostructures of atomically thin TMDs. This technique combines ultrafast optics and efficient data acquisition with advanced data processing and analysis to get a complete picture of the photoresponse over all relevant experimental parameters. In this section, we discuss the optics (Section S1.1), data acquisition (Section S1.2), and data analysis concepts (Sections S1.3 to S1.5) involved in this technique.

## S1.1. Optics

MPDPM involves space-time resolved photocurrent measurements, which require an optical probe that can vary controllably in space and time. Our optics combine the techniques of scanning beam photocurrent and reflectance microscopy with ultrafast optical pump-probe measurements (*22-27, 31, 32*). A schematic of the optical system is shown in Figure S1A. The light source is a MIRA 900 OPO ultrafast pulsed laser which generates 150 fs pulses with controllable wavelength from 1150 nm to 1550 nm at a 75 MHz repetition rate. The output of the laser is split into two paths by a 50/50 beamsplitter and a translation stage is used to controllably introduce a path length difference. The two beams are then combined and the path length difference splits a single laser pulse into two sub-pulses separated by a time delay, $\Delta t$.

The recombined beam is then fed into scanning beam optics which consists of rotating mirrors and a system of two lenses that focus the beam onto the back of an objective lens. The objective lens is set at the focal length of the second lens such that, as the scanning mirror rotates the beam is still focused onto the same position on the back of the objective, but arriving at the different angle. The objective lens focuses the light down into a diffraction limited beamspot where the position of the beamspot depends on the incident angle. As the scanning mirror rotates, the beamspot moves over a wide range of the sample without aberration, allowing for quick high-resolution scanning.

Figure S1B details our specific scanning optics and the customized Janis Research ST-3T-2 optical cryostat that we use in our experiments. The sample sits at vacuum on a sample stage, which can vary the temperature from 4 K to 420 K. The sample stage is in the center of a 3 Tesla superconducting magnet. Devices are electronically probed using four probe needles which contact conductive pads on quartz chip carriers that are wire-bonded to fabricated gold contacts on the sample. We amplify the electrical signal and measure the current resulting from the incident laser light, or photocurrent. We also measure the reflectance of the sample by measuring the intensity of the light that is reflected from the sample with a near-infrared photodiode.

To fully enclose our focusing optics inside the vacuum chamber, we use a Gradient Index of Refraction (GRIN) lens as an objective. A GRIN lens is a single small cylinder of glass in which the index of refraction is continuously varied along the radial and axial directions. Since it lacks the many interfaces of a conventional objective, a GRIN lens does not disperse the laser pulses as dramatically as a traditional objective lens. Figure S2A shows the autocorrelation of the reflected intensity due to two overlapping laser pulses, taken by scanning the delay stage near $\Delta t = 0$. The width of the autocorrelation pattern is approximately three times the pulse width, our autocorrelation is about 570 fs wide indicating that our pulses are approximately 190 fs long at the sample, very close to the laser specification of 150 fs. The low pulse dispersion allows us to measure short timescales accurately and gives a high peak pulse intensity.

When properly aligned the power throughput of the GRIN lens is near unity, allowing us to focus tens of milliwatts of power onto the sample. The laser power is calibrated under the assumption of ideal GRIN power throughput, and the process of carefully aligning the optic onto the sample under vacuum introduces systematic uncertainty into the laser power. Though it is difficult to account for this uncertainty exactly, we estimate that it is less than 10%. Thus, all power measurements shown in this work have a systematic uncertainty of 10%.

The resolution of a microscope is limited by the diffraction limit, which gives a minimum resolvable feature size of $d_{min} = \lambda/2N_A$, where $\lambda$ is the wavelength of the light and $N_A$ is the numerical aperture of the lens (*33*). This limits the optical probing of samples that may be a few microns or less in size. In principle, when focused, the laser beamspot spatial profile is an Airy disk, which can be approximated as a Gaussian function, which we take as our point spread function. Figure S2B shows the measured photoresponse of an absorber smaller than 1 $\mu$m using a wavelength of 1200 nm. The data is fit well by a Gaussian function (solid line) with full width at half maximum of 1.67 $\mu$m, indicating that our system is near the diffraction limit.

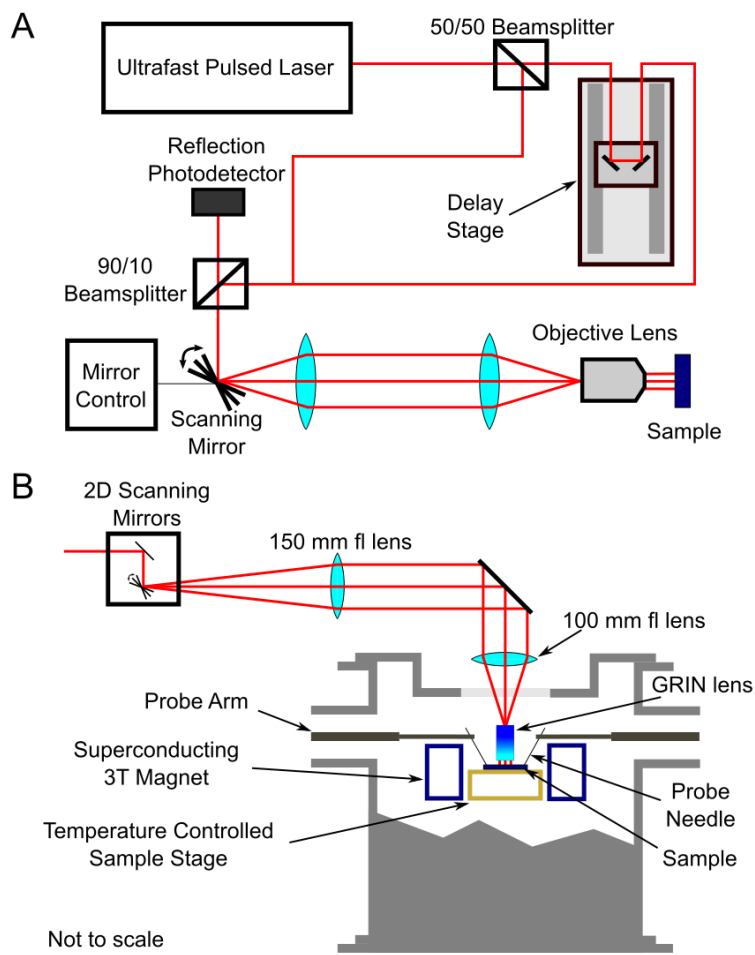

**Fig. S1.**

*Schematic diagram of the MPDPM optics.* (**A**) A conceptual picture of the optical setup showing all the major optical components. (**B**) A cross sectional diagram of the optical setup and optical cryostat detailing the optics coupling into the GRIN lens.

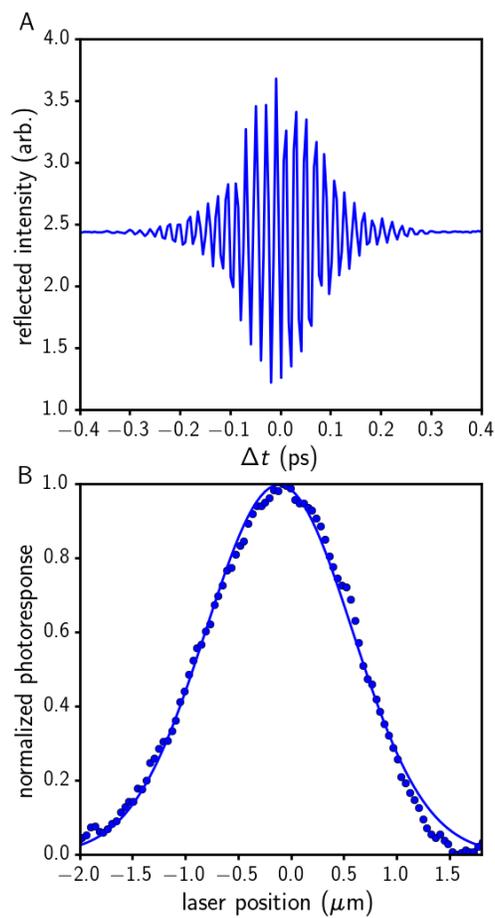

**Fig. S2.**

*Characterization of MPDPM pulsed laser optics.* (**A**) Two pulse autocorrelation versus delay between two subsequent pulses, $\Delta t$. (**B**) Measured photoresponse of an absorber smaller than the diffraction limit. Solid line is a fit to a Gaussian function with a full width at half max of $1.67\mu$m.

**S1.2. MPDPM Data Acquisition**

The main innovation of the MPDPM technique is rapid and efficient data acquisition. We developed an integrated Data Acquisition (DAQ) program using a set of python modules that interface with equipment drivers and control all hardware components simultaneously. The experimental setup was designed to allow the maximum amount of automation possible, so that data could be acquired rapidly, systematically and repeatably. Our experimental setup can scan a beam in two dimensions while applying voltages to the sample under various optical conditions. In addition, the optical cryostat that contains our samples can control the temperature of the sample and apply a magnetic field. Each of these components requires specialized hardware, which we designed and selected to allow for full automation.

The DAQ system involves a complex system of data and feedback shown schematically in Figure S3. The main hardware components of the optics and controllers are shown in the upper left. These components are controlled with feedback to the DAQ software, which is represented in the lower left. From the user interface, any of the hardware components can be changed or scanned, varying some output over a given range. If one of the components is set to scan, the rest will be held to constant values. The most common scan is a rectangular scan of the 2D scanning mirrors, which moves the laser beam spatially over the surface of the sample. The software is designed to consistently and time-efficiently take scans, such that many images can be rapidly acquired as a function of other variables.

There are two types of data collected. First, analog voltage data, or signal, comes out of the optical cryostat and two InGaAs photodiodes. The signal is appropriately amplified, and then acquired through a National Instruments PCIe-6323 Data Acquisition Card (Figure S3 upper right). The DAQ software controls the NI card through the commercial NI-DAQmx drivers, which return a time series of data digitized at a frequency set by the card clock. The second form of data is the hardware control data, from the hardware feedback and control driver systems, which determine what is happening to the samples during the experiment. For the data to be usable these two types of data must be correlated. For example, if the laser beam is scanned over the surface of the sample, the position is controlled by the hardware drivers. The resulting signal comes in through the card as a function of time, the DAQ software correlates the output of the scanning mirror controllers with the signal coming in a short time later, relating the signal to the physical location of the laser beamspot on the sample. Data from the card is then binned into a two-dimensional array with a parameter (in this example laser beamspot position) varying along each axis.

Runs are the discrete unit of data for our experiments, with each sampling a particular volume of parameter space. When written to disk the data is packaged together into one data set called a "run". In addition to the data, each run saves all possible control parameters, 125 in total, of the hardware and software to ensure consistency and repeatability. Each run is assigned a unique run number and the files for that run are saved to disk in a data archive. Each run can take several minutes to several hours, and collect a few dozen kilobytes to tens of megabytes of raw data. Typically, a sample will require hundreds of runs to fully examine and even then, the total possible parameter space is often too large to examine completely, requiring tradeoffs to be made. For example, lower resolution scans are often faster, therefore we may choose to take more lower resolution scans to cover more parameter space or to take fewer but higher resolution scans to observe an effect more accurately but over less parameter space. These tradeoffs must be made carefully, with knowledge of how data analysis will eventually extract results, to avoid missing data.

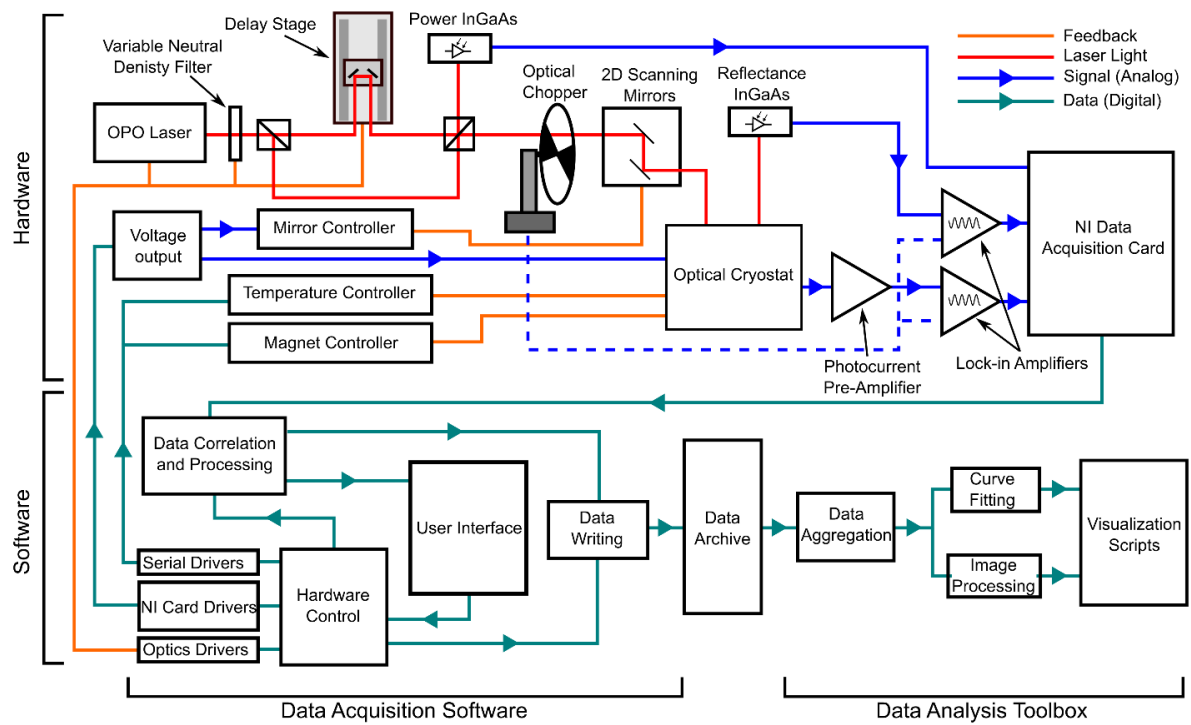

**Fig. S3.**

*Experimental data flow for rapidly acquired MPDPM imaging.* Schematic of the flow of data between hardware and software components as well as the feedback involved in controlling the experiment.

### S1.3. Data Structure in MPDPM

The MPDPM technique generates multi-dimensional data sets that can be understood as cubes or hypercubes of data. The data acquisition software generates data by scanning parameters, such as laser position, and measuring the response. The resulting data is a two-dimensional image or "data plane" that describes behavior of the sample as a function of space or other parameters. The experimental setup can scan consistently over the sample many times in succession, varying some experimental variable between scans. This gives a series of data planes which are processed into a three-dimensional "data cube" which depends on three independent variables, two spatial variables and a third "vertical" axis that may be laser power, optical delay, applied voltage etc. Once we can consistently construct a data cube we can vary any of the other experimental parameters we are controlling, such as voltage, wavelength or magnetic field, taking many data cubes and constructing a "data hypercube" a data set with four independent variables. Theoretically, we could continue this to an arbitrary number of independent variables, taking exponentially more data with each dimension added.

MPDPM allows access to four or more independent variables, yet most data processing and human intuition is suited for only one or two independent variables. Our approach is to use data fitting and image processing to reduce the number of independent variables to the two spatial dimensions. First, fit the data to a phenomenological law. Then use a fitting parameter, which contains information about the underlying physics, as a metric for the behavior of the data. This reduces the number of independent variables to three, turning a data hypercube back into a data cube. From this reduced data cube, we can identify physically important features using image analysis (discussed in section S1.5) and reduce the number of independent variables to two by projecting these features onto the spatial axes. Overall, our approach is to take a large multivariable data set and then use fitting and various forms of analysis to condense it back to a manageable amount of data.

### S1.4. Data Processing of the MPDPM images

We use a set of custom python modules, together forming a "toolbox" to handle data runs in a systematic manner. The lower right section of Figure S3 shows the main functions of the toolbox. Given a run number, the code retrieves the and any relevant calibration data and returns the calibrated data along with all the experimental parameters. The next step is image processing and curve fitting. Image processing involves combining two dimensional images into a larger data set, such as constructing a three-dimensional data cube from a series of data planes. In that case the image processing must account for the physical drift in the images. To correct for drift the software computes the two-dimensional autocorrelation function $g(x, y)$, of two subsequent images $I_1$ and $I_2$, defined by the convolution integral:

$$g(x, y) = \int_{-\infty}^{\infty} \int_{-\infty}^{\infty} I_1(t_x, t_y) I_2(x - t_x, y - t_y) dt_x dt_y \tag{1}$$

The coordinates of the maximum of the autocorrelation function, $(\Delta x, \Delta y) = Max[g(x, y)]$, gives the drift between $I_1$ and $I_2$. The autocorrelation maximum is located using a customized peak finding algorithm, optimized for this type of data. Then one of the images is shifted by $(\Delta x, \Delta y)$ thus eliminating the drift, and allowing points on the two subsequent images to be considered together. This is done for all images in sequence so that data points are spatially correlated between all images.

Curve fitting involves taking correlated data points and fitting to a phenomenological law using a non-linear least squares fitting algorithm. The phenomenological law can be any function that parameterizes the data, most commonly we use equations $I \propto P^{\gamma}$ and $I \propto e^{-t/\tau}$. Then phenomenological parameters, such as $\gamma$ and $\tau$ are extracted. Due to noise in the data some of the curve fits will return extreme or meaningless results. This data must be identified and eliminated and several functions were developed to filter the bad results out based on various noise sources. Depending on type of run, image processing and curve fitting can occur independently, or in series. In addition, image processing can be used to correlate curve fit data between multiple runs to construct data hypercubes.

### S1.5. Image Analysis in MPDPM

Image analysis is an important tool for MPDPM. Data cubes and hypercubes are usually composed of many two dimensional spatial scans, image analysis allows us to handle these data sets by identifying key features in images and looking at how those key features depend on experimental parameters. The data used in main text Figure 2 is a good example; the images contain a physically interesting "ring" feature in the center of the sample, and before we can analyze the ring we must quantify it using image analysis. The entire data set used in main text Figure 2 is shown in Figure S4A.

For each of the photocurrent images in Figure S4 we need to determine if there is a ring in the center, and, if so, measure the area of the ring feature. Maps of the magnitude of the gradient vector are shown for the data in Figure S4B. We see that where there is a ring in the photocurrent it is enclosed by a contour where the gradient is zero, i.e. where the photocurrent stops increasing and starts to decrease. To pick out this contour we use an iterative algorithm, which at the edges of the photocurrent and at each iteration uses the gradient vector as a "force" to accelerate the points of the contour. This will cause the contour to "climb" the photocurrent, shrinking inwards as it moves up along the gradient. If there is no dip feature the contour will shrink until it has zero size at the maximum. If there is a dip feature, then the contour will get stuck along the edge of the dip, because the gradient is zero at that point. Once the contour converges either to zero or to a stable finite size, the algorithm returns the contour which is used to generate main text Figure 2C. An example of this contour is shown on the bottom right maps of Figure S4. This algorithm is robust, as noise in the images has little effect on the returned contour.

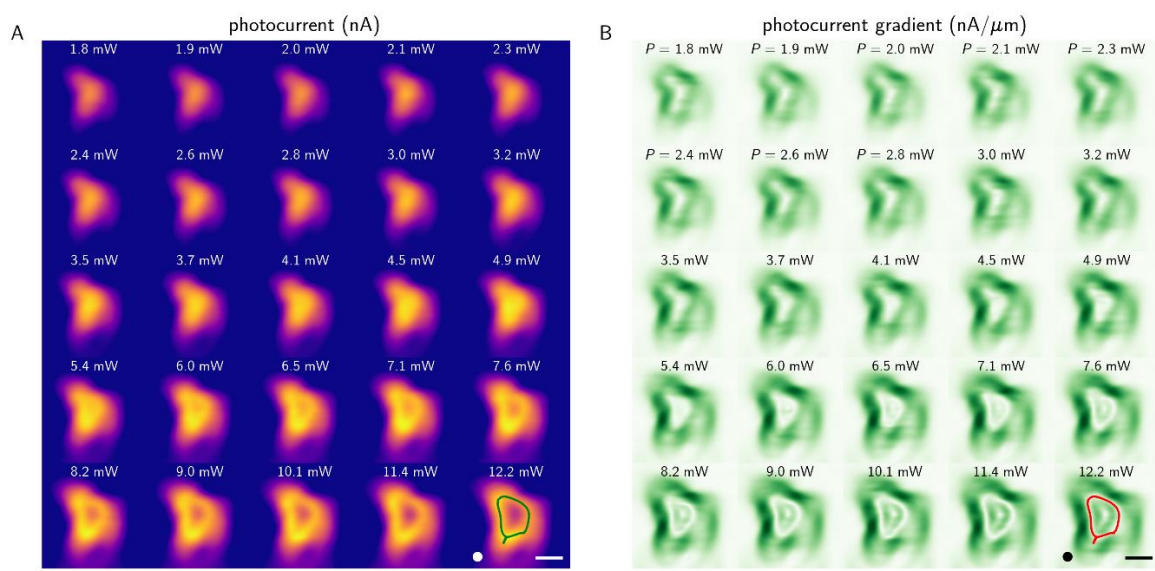

**Fig. S4.**

*Typical MPDPM photocurrent maps from a single parameter measurement.* (**A**) Images of the data set of photocurrent scans as a function of power at $\Delta t = 0.2$ ps and $V_i = 0$. Used to generate the data shown in main text figure 2. (**B**) The magnitude of the gradient of photocurrent from the images in A. At high power a contour of $|\nabla I| \approx 0$ occurs at the maximum of the photocurrent ring. This contour is identified algorithmically, and an example of contour is plotted in the solid red line in the lower right. In the lower right image of both A and B the circle represents the 1.67 $\mu$m beamspot FWHM and the scale bar is 5 $\mu$m.

## S2. Fabrication and Characterization of the Graphene-MoTe$_2$-Graphene Photocells

Ultrathin MoTe$_2$ is an ideal semiconductor material for the exploration of strongly interacting electron-hole pairs and electron-hole condensation. The electron-hole liquid phase requires the existence of electron-hole pairs with sufficiently long radiative lifetime to form highly dense plasma under optical illumination (8). Electron-hole condensation has been observed in only a few materials including silicon and germanium (7,8, 34,35), both of which exhibit indirect band gaps that give rise to very low photoluminescence efficiency. In the ultrathin form, MoTe$_2$ exhibits an indirect band gap of approximately 1.0 eV (19, 36), similar to silicon. Among TMD materials, MoTe$_2$ displays the weakest photoluminescence, with easily measured response occurring only in the monolayer and dropping off dramatically as the thickness increases (19). In contrast, MoTe$_2$ exhibits very high electronic photoresponse. The highest photoresponse observed to date is in MoTe$_2$-graphene heterostructures (15,17,18), with values as high as 87 AW$^{-1}$ (16), outperforming most of the TMD materials studied to date, and demonstrating the potential for MoTe$_2$ in a variety of optoelectronics applications.

We fabricated graphene-MoTe$_2$-graphene (G-MoTe$_2$-G) heterostructures in order to investigate the characteristics of the interlayer photocurrent. Our fabrication technique successfully produces well-ordered heterostructures that can be characterized by Raman spectroscopy and AFM to identify basic material properties as well as give an indication of the relative thicknesses of each layer. In this section, we provide details of this fabrication process (Section S2.1) and analysis of the characterization data (Section S2.2).

## S2.1 Synthesis of Graphene-MoTe$_2$-Graphene Heterostructure Photocells

We obtain bulk graphite from Covalent Materials Corporation and MoTe$_2$ samples from 2D Semiconductors Inc. Mechanical exfoliation of bulk samples onto Si/SiO$_2$ (290 nm oxide) substrates yields atomically thin flakes of graphene and MoTe$_2$. Prior to exfoliation, we clean our Si substrates in acetone for approximately 20 minutes.

We employ a dry transfer process, based on a technique developed by Andres Castellanos-Gomez, et al. (37), to construct heterostructures out of these constituent flakes. Figure S5 shows the details of the transfer method. We utilize a custom-built transfer microscope consisting of two essential components: a stage that holds the sample and a cantilever. A silicon substrate (containing a desired flake for transfer) attaches to the stage with carbon tape (Figure S5A). A stamp composed of a piece of polydimethylsiloxane (PDMS) placed onto a standard glass slide with a thin layer of polypropylene carbonate (PPC) spin-coated on top, attaches to the cantilever through vacuum suction. Initially, the stamp lowers to contact a graphene flake (Figure S5B) and the stage heats up to 40°C before cooling back down to room temperature. The cantilever then detaches rapidly from the stage, allowing the graphene to transfer from the substrate to the stamp (Figure S5C). Next, a new Si substrate containing a MoTe$_2$ flake attaches to the stage and the cantilever lowers to allow for alignment of the MoTe$_2$ directly underneath the graphene flake on the stamp. The stamp then contacts the substrate and undergoes the same heating procedure to transfer the MoTe$_2$ flake to the stamp.

Finally, we place a Si substrate with a thicker graphene flake onto the stage and once again lower the cantilever, aligning the graphene flake on the substrate with the existing heterostructure on the stamp and bringing the two into contact. This time, the stage heats to 80°C, allowing the PPC layer to melt. Then, by slowly lifting the cantilever, the PPC separates from the PDMS and bonds to the sample, leaving the completed 3-layer heterostructure on the substrate (Figure S5F).

Afterwards, we clean the sample in acetone for at least an hour to ensure removal of the sacrificial PPC layer.

We electrically contact our heterostructures using standard electron-beam lithography followed by electron-beam evaporation of 4 nm of Ti and 120 nm of Au. A Leo SUPRA 55 and Temescal BJD 1800 perform electron-beam lithography and evaporation procedures, respectively.

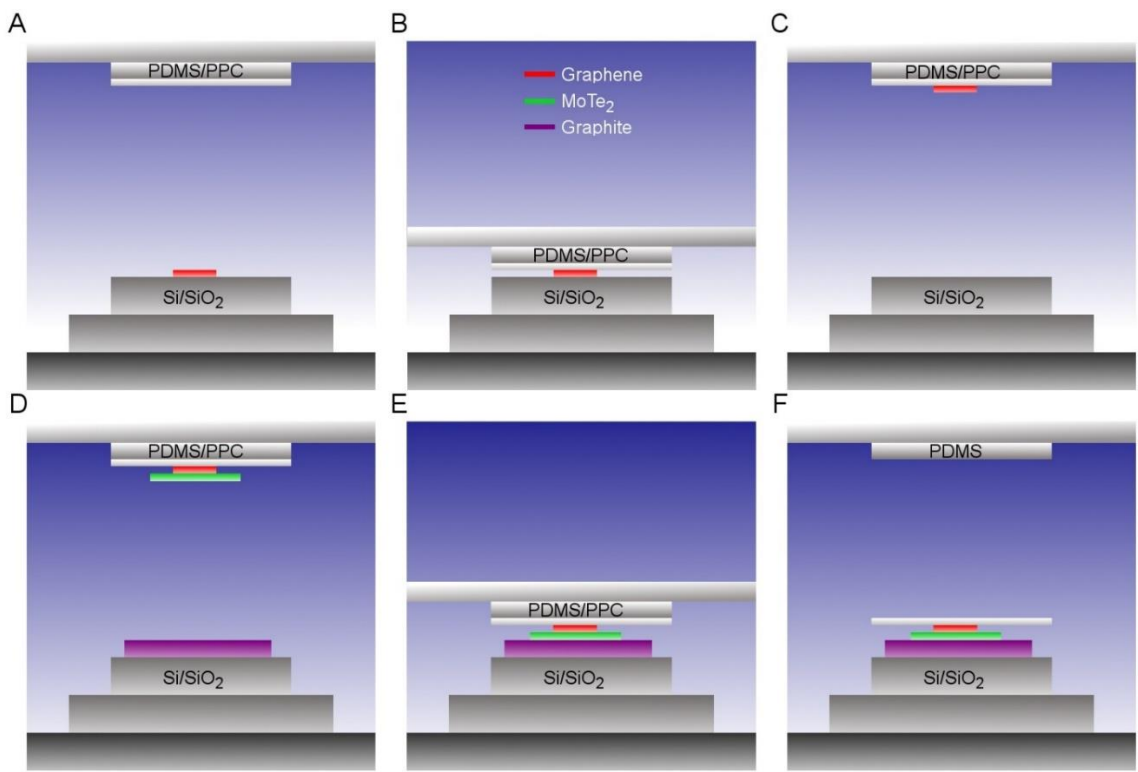

**Fig. S5.**

*Diagram of dry transfer method for assembly of heterostructures.* (**A**) Initially, a graphene flake is placed on the sample stage and a stamp is made consisting of a glass slide, PDMS, and PPC. (**B**) The stamp is lowered until it encounters the graphene on the sample stage. (**C**) After heating the stage to 40˚C and allowing it to cool back to room temperature, the stamp is quickly removed from the stage, picking up the graphene flake. (**D**) The same procedure is used to pick up the next layer of MoTe$_2$, and then the final layer of graphene is placed on the sample stage. (**E**) The flakes are aligned on top of each other and the stamp is lowered until it contacts the stage. (**F**) The stage is heated to 80˚C to melt the PPC layer, and the stamp is raised slowly to allow the PPC to separate from the PDMS, leaving the completed heterostructure on the desired substrate. The PPC layer is then removed in an acetone bath.

## S2.2. Raman Spectroscopy and Atomic Force Microscopy (AFM)

Raman spectroscopy and AFM are used to determine the relative layer thicknesses in each sample. We analyze two devices, the one discussed in the main text (Device 1) as well as a secondary device (Device 2). Spectroscopic characterization was performed using a Horiba LabRam in the backscattering configuration with a 20 mW of laser power at a wavelength of 532 nm, a 100x objective, and a 1800 grooves/mm grating. Figure S6 and Figure S7 show Raman spectroscopy measurements for Device 1 and Device 2, respectively. Figures S6A and S7A show optical images of both devices, as well as an outline of the heterostructure region. Figure S6B displays the Raman spectrum for the top graphene layer in Device 1. We observe the characteristic G and 2D peaks at 1576 $cm^{-1}$ and 2716 $cm^{-1}$, respectively. The ratio of the 2D peak to the G peak can be used to roughly estimate the layer thickness, indicating a thickness of roughly 5-8 layers (*38*). Similarly, Figure S7B shows the Raman spectrum for the top layer of graphene Device 2, where the G and 2D peaks occur at 1578 $cm^{-1}$ and 2714 $cm^{-1}$, respectively, and it appears to be slightly thinner than the graphene in Device 1, approximately 4-5 layers.

Figures S6C and S7C present Raman spectra for the $MoTe_2$ layers in each device. In Figure S6C, the characteristic peaks associated with $MoTe_2$ in Device 1 can be clearly identified. (*19, 39-41*). We see the $A_{1g}$ peak at 168 $cm^{-1}$, the $E^1_{2g}$ peak at 229 $cm^{-1}$, the $B^1_{2g}$ peak at 285 $cm^{-1}$, and the Si peak at 517 $cm^{-1}$. Likewise, Figure S7C shows the characteristic $A_{1g}$ peak at 169 $cm^{-1}$, the $E^1_{2g}$ peak at 230 $cm^{-1}$, and the $B^1_{2g}$ peak at 287 $cm^{-1}$, along with the Si peak at 517 $cm^{-1}$ for Device 2. In $MoTe_2$, layer thickness can be roughly estimated from the ratio of the $B^1_{2g}$ to the $E^1_{2g}$ peaks (*19, 40*). The Raman data suggests the presence of both 2H and 1T' phases of $MoTe_2$, as reported by Cho, S. et al. (*42*).

Figures S6D and S7D give the Raman spectra for the bottom graphene layer in each device. Figure S6D displays characteristic G and 2D peaks at 1576 $cm^{-1}$ and 2717 $cm^{-1}$, respectively, whereas Figure S7D presents G and 2D peaks at 1579 $cm^{-1}$ and 2716 $cm^{-1}$, respectively. Both plots reveal a shoulder near the 2D peak at 2687 $cm^{-1}$, which is indicative of thicker graphene (*38*).

We perform AFM measurements using a Digital Instruments Nanoscope IV with a silicon cantilever in tapping mode to confirm the layer thicknesses of each device. For Device 1, the top graphene, $MoTe_2$, and bottom graphene layers display thicknesses of 5.88 nm, 8.96 nm, and 13.8 nm, respectively. For Device 2, the top graphene, $MoTe_2$, and bottom graphene layers show thicknesses of 3.19 nm, 13.6 nm, and 6.12 nm, respectively. These results agree with expectations from the Raman spectroscopy measurements.

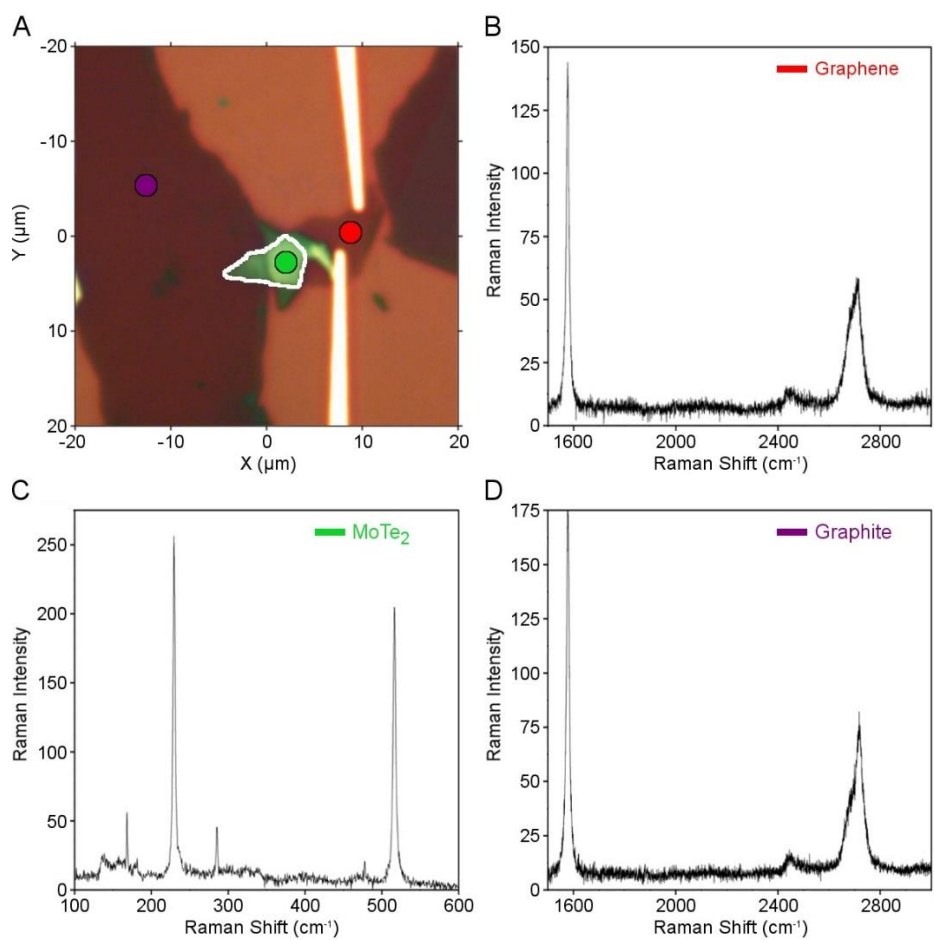

**Fig. S6.**
*Raman Spectroscopy of Device 1.* (**A**) Optical image of Device 1 with graphene, MoTe₂, and graphene layers indicated with color-coded markers. Blue outline indicates region of 3-layer overlap. (**B**), (**C**), (**D**) Raman spectra of the top graphene, middle MoTe₂, and bottom graphene layers, respectively, taken at the position of the corresponding markers in (A).

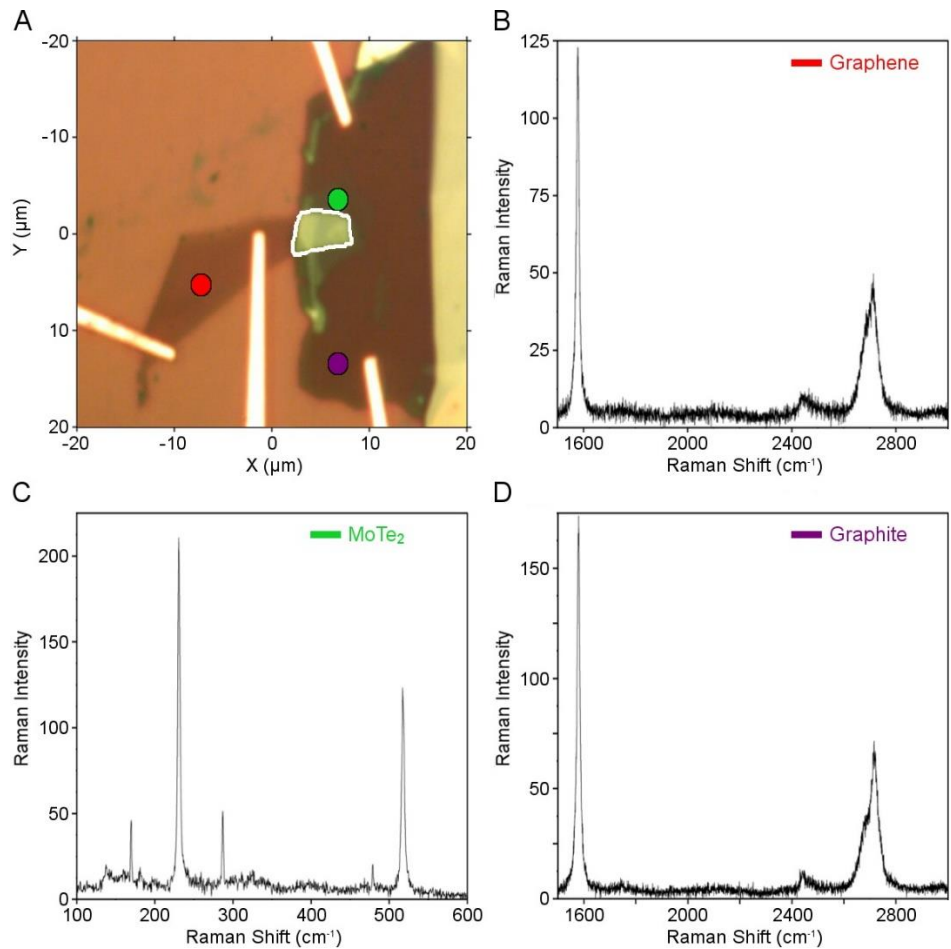

**Fig. S7.**

*Raman Spectroscopy of Device 2.* (A) Optical image of Device 2 with graphene, MoTe2, and graphene layers indicated with color-coded markers. Blue outline indicates region of 3-layer overlap. (B), (C), (D) Raman spectra of the top graphene, middle MoTe2, and bottom graphene layers, respectively, taken at the position of the corresponding markers in (A).

**S3: Optoelectronic Characterization of the Graphene-MoTe₂-Graphene Photocells**

In this section, we discuss the dependence of the interlayer photocurrent-voltage characteristics on incident laser power, temperature, and wavelength. We first discuss the ordinary photocurrent-voltage characteristics in the absence of the electron-hole liquid and their dependence on laser power at low power. We then present the temperature and wavelength dependence of the interlayer photocurrent-voltage characteristics.

**Section 3.1. Photoresponse below the threshold for electron-hole condensation**

We first characterize the photoresponse at low interlayer voltage, in the absence of the electron-hole condensate. Figure S8 compares a typical spatially resolved photocurrent map to an optical image of the heterostructure photocell device (with an outline of the heterostructure region). As expected, the photocurrent is generated primarily in the heterostructure overlap region. We measure the change in photocurrent as we vary the interlayer bias voltage and incident laser power, generating the characteristics shown in Figure S8C. The interlayer photocurrent increases linearly with low bias voltages, as seen in previous literature (*43*). We observe an offset in the zero-crossing of the photocurrent that depends weakly on laser power, indicating the presence of an open-circuit voltage, which has been reported previously in G-MoTe₂-G photocells as well as in other TMDs (*15,17,44*). From Figure S8C, we extract an open circuit voltage of $\phi_0 = -41$ mV for Device 1.

We observe that the photoresponse of the G-MoTe₂-G photocell is strongly sensitive to sample temperature, and exhibits behavior consistent with a very lightly doped, indirect gap semiconductor. Figure S9A shows a series of current-voltage characteristics as a function of temperature. Higher temperature yields significantly more photocurrent, as observed by Lin et al. (*45*). The photocurrent diminishes with lower temperature since there is less thermal energy to excite carriers from the valence to the conduction bands. This behavior suggests that the MoTe₂ is nearly intrinsic, reaching strongly insulating behavior at low temperatures. At higher bias voltages, the photocurrent-voltage characteristics become nonlinear, which is also consistent with previous work (*18*). The nonlinear photocurrent response may be due to the presence of Schottky barriers between the MoTe₂ and graphene layers, as reported by Zhang, K. et al. (*15*). In the measurements shown in the main text, we measure only at low interlayer voltage, where the photocurrent-voltage characteristics are linear.

Wavelength-dependent measurements indicate that the highly efficient photoresponse in the G-MoTe2-G photocell occurs only when the incident photon energy $E_{PH}$ matches the bang gap energy of MoTe₂. Figure S9B shows a series of photocurrent-voltage characteristics as a function of wavelength. The photocurrent increases as the wavelength decreases, reaching at maximum photoresponse at $\lambda = 1200$ nm. MoTe₂ exhibits a high valance band splitting of roughly 300 meV (*46*) and two primary absorption peaks corresponding to the A and B excitons at approximately 1.08 eV and 1.425 eV (*21*). The strong photoresponse observed here is consistent with the onset to photon absorption at the A exciton band edge in MoTe₂, which we observe at $\lambda = 1200$ nm ($E_{PH} = 1.03$ eV).

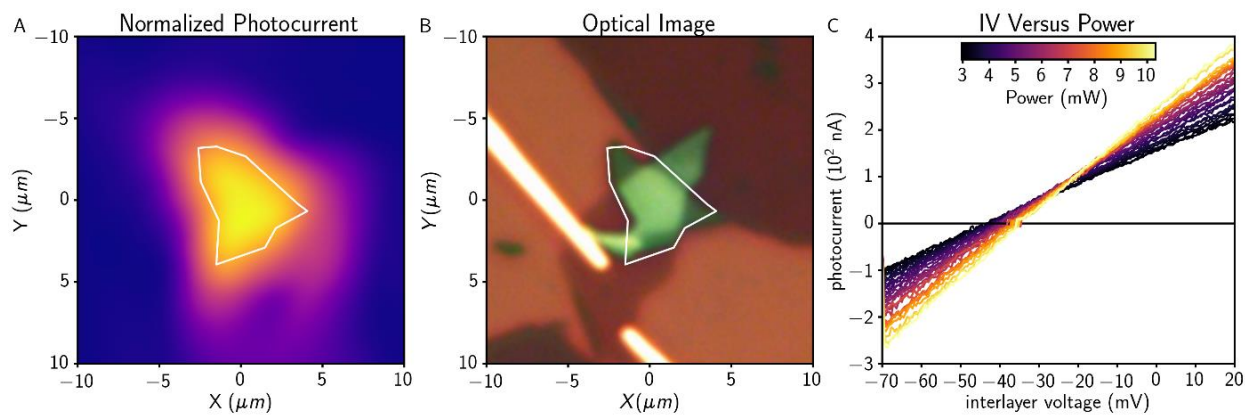

**Fig. S8.**

*Optoelectronic I-V characteristics.* (**A**) Optical image of device with heterostructure region again outlined in white. (**B**) Photocurrent map of region shown in (A) with overlap region again outlined in white. (**C**) I-V curves for various incident laser intensities. Data taken at room temperature and 1200 nm wavelength.

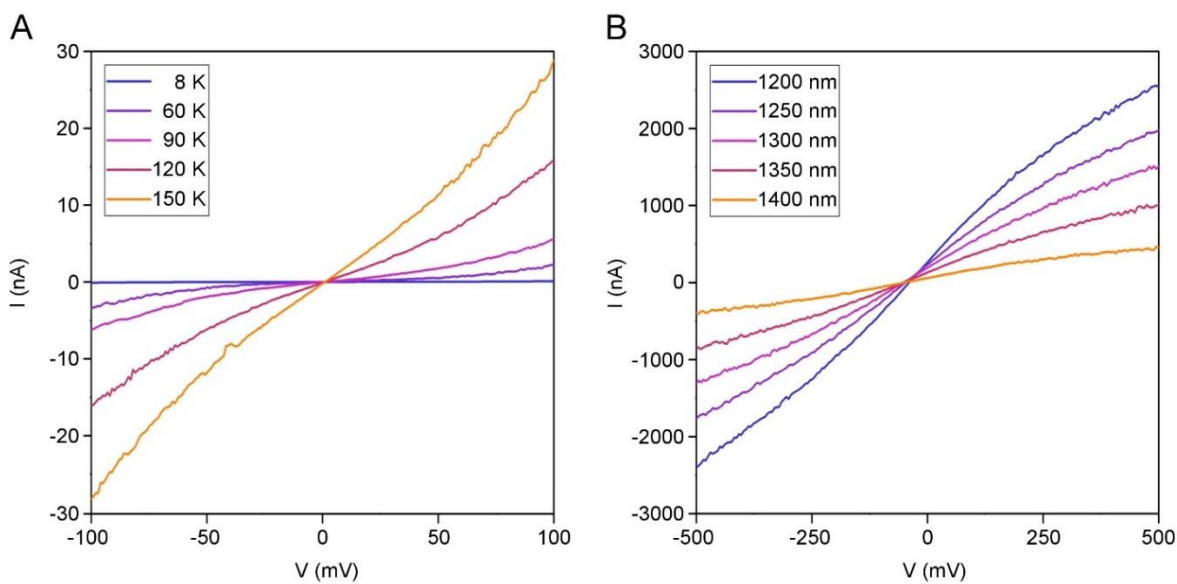

**Fig. S9.**

*Temperature and wavelength dependence of interlayer current-voltage characteristics.* (**A**) I-V curves at various temperatures. Power = 23.0 mW, Wavelength = 1200nm. (**B**) I-V curves at various incident laser wavelengths. Power = 4.80 mW, Temperature = 296 K.

**Section 3.2. Photoresponse at the threshold for electron-hole condensation**

Above the critical threshold for electron-hole condensation the photoresponse exhibits significant changes in behavior as a function of power and $\Delta t$. Figure S10 shows the power dependence using line cuts through the center of the sample, as a function of power, to get a detailed characterization of the power dependence. The photocurrent is suppressed in the center of the sample at high values of power and low values of $\Delta t$ consistent with the electron-hole liquid model. Figure S10F shows the power dependence at the center of the line cut, which shows strong suppression of photocurrent. Beyond looking at the photocurrent from any individual point we also examine the spatially integrated photocurrent, shown in Figure S10G. At short $\Delta t$, the integrated photocurrent flattens out and is approximatly constant after the critical point.

Figure S11 shows similar data to Figure S10, instead looking at the differential reflectance ($\Delta R/R$). The reflectance exhibits a common form that can be fit to a power law with a negative exponent. Looking at the residuals to the fit, a weak bump in the data can be observed which is consistent with a metal-insulator phase transition. However, the signal is weak and does not give as clear an indicator as the photocurrent data.

Figure S12 shows a detailed measurement of the voltage dependence, from the same data set as main text Figure 4A. The size of the photocurrent dip, $\ell$, is observed to decrease as a function of increasing interlayer voltage. The square of $\ell$ decreases linearly with respect to voltage, as expected as $\ell^2$ measures the area of the photocurrent ring.

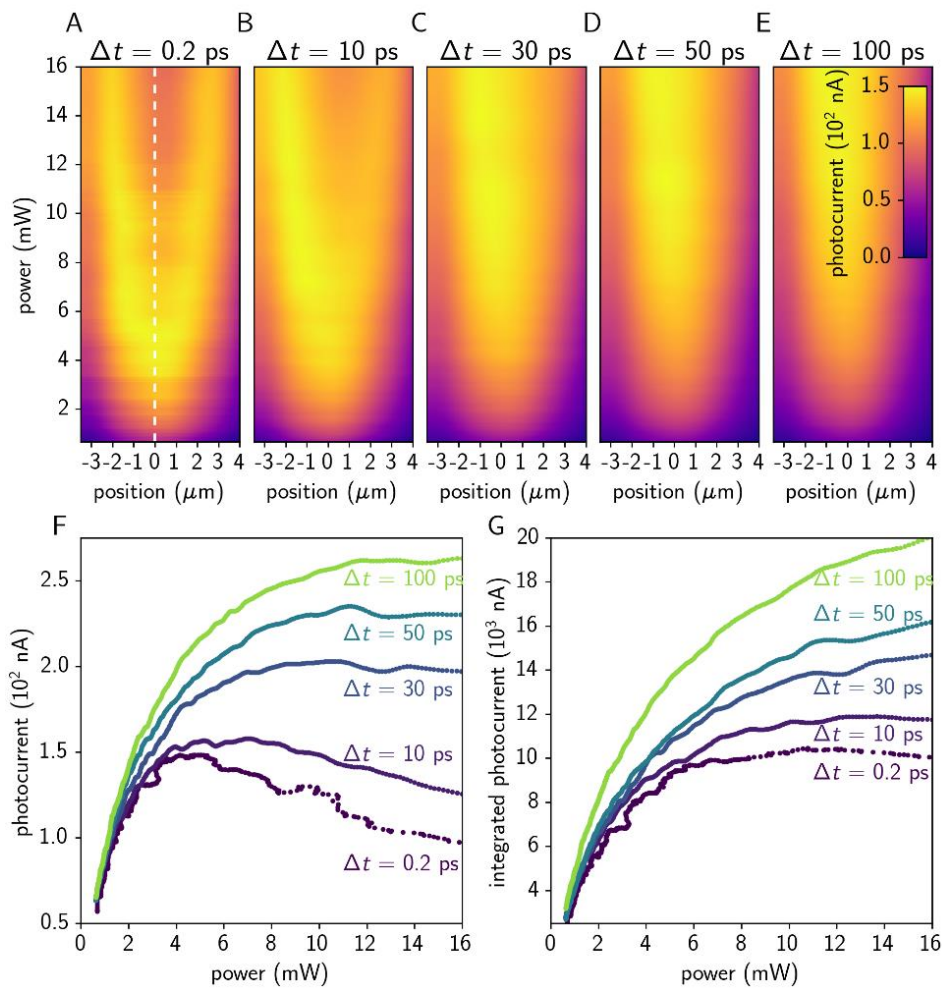

**Fig. S10.**

*Photocurrent line scans across the threshold of electron-hole condensation.* (**A-E**) Photocurrent line cuts, moving the laser in a line across the center of the sample as a function of power, for various values of $\Delta t$. (**F**) The photocurrent versus power at the center of the sample, (along the dashed line in a). (**G**) The photocurrent integrated along the line scan as a function of power.

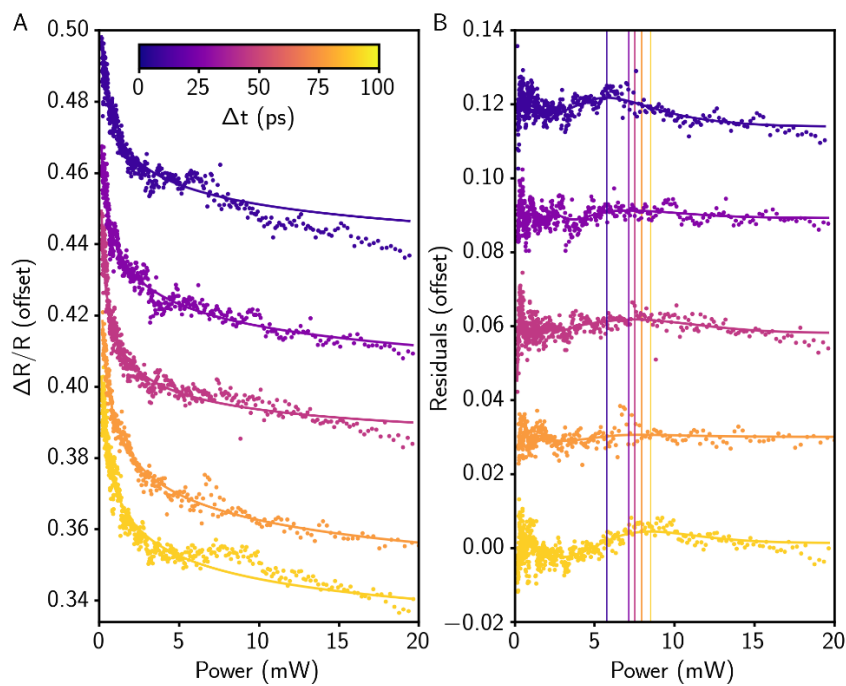

**Fig. S11.**

*Reflectance measurements.* (**A**) ΔR/R as a function of power for various time delays, solid lines show a power law fit. (**B**) ΔR/R fit residuals as a function of power, showing weak peaks. The center of each peak is denoted by vertical lines of corresponding color to the dataset.

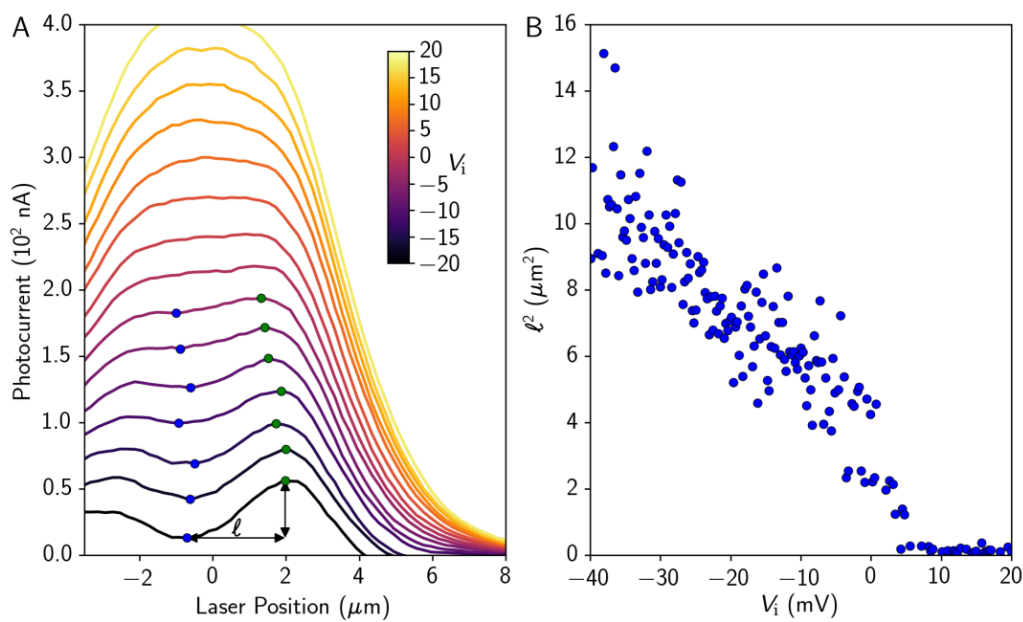

**Fig. S12.**

*Detailed voltage dependence.* (**A**) Line cuts of photocurrent as a function of interlayer voltage, $V_i$. The quantity $\ell$ is used the track the "dip" feature in the photocurrent, measuring the distance from the local minimum (blue dots) to the shoulder (green dots). (**B**) The square of $\ell$ versus voltage.

## S4: Supporting Calculations

### S4.1 Estimation of the Mean Distance Between Electron-Hole Pairs at $P_C$

To calculate the average distance between the electron-hole pairs at the critical point of the phase transition, we must find the density at the critical power $P_C$. The number of photons incident on the sample by a single pulse of the laser is given by $N = E_{pulse}(P)/E_{photon}(\lambda) = \lambda P/hcR$. Where $R = 75$ MHz is the repetition rate of the laser, $\lambda = 1200$ nm is the laser wavelength, and $P$ is the power of the laser beam. The density of electrons and holes generated by a single pulse is $n = 2AN/V$. Where $A$ is the percentage of light absorbed by the sample, and $V$ is the total volume containing the electron-hole population. The average distance between charge carriers, $d_{avg}$, is then given by $\frac{4}{3}\pi d_{avg}^3 = \frac{1}{n}$ which gives the average spacing between electron-hole pairs

$$d_{avg} = \sqrt[3]{\frac{3}{8\pi}\frac{V}{AN}} \approx 1.4 \times 10^{-4} \sqrt[3]{\frac{V}{AP}} \qquad (2)$$

To estimate $d_{avg}$, we must estimate the volume $V$ in which the electron-hole population undergoes condensation. This volume depends on how fast the electron hole liquid forms, thus we can only place bounds on it. For the lower bound, the phase transition occurs quickly relative to the timescale of diffusion and the volume is a cylinder with diameter equal to the FWHM of the beamspot and height equal to the sample thickness. Therefore $V_{min} = \pi(1.67\ \mu m)^2(9 nm) = 0.08\ \mu m^3$ which at the critical power of $P_c = 6$ mW gives an average distance of $d_{avg} = 0.71$ nm (and corresponding density of $n = 0.66\ nm^{-3}$), assuming 10% absorption in MoTe$_2$ (*36*). For the upper bound, carriers fully diffuse into the total volume of the sample prior to condensation. The area of the sample can be estimated from Figure S8 which gives an area of $66\ \mu m^2$, therefore $V_{max} = 5.94\ \mu m^3$ which gives an average distance of $d_{avg} = 3$ nm (and corresponding density of $n = 0.09\ nm^{-3}$). Thus, we conclude that at the critical point, charge carriers are separated by an average of 1-3 nm. This value is similar to values of the exciton Bohr radius determined through magneto-optical measurements of MoTe$_2$ (*28*).

### S4.2 Spatially-Resolved Photocurrent in the Electron-Hole Liquid Phase

To understand the spatial distribution of the photocurrent we need to consider what happens when a laser pulse illuminates the sample. At the surface of the sample the laser is a diffraction limited beamspot which can be approximated as a Gaussian with full width at half max equal to 1.67 microns. Therefore, the spatial profile is given by $P(x) = P_0 e^{-\frac{x^2}{2\sigma^2}}$ in one dimension. Below the critical threshold, $P_C$, the photoresponse obeys a power law $I \propto P^\gamma$. Thus for $P_0 \leq P_{th}$ the observed photocurrent is given by

$$I(x) = \int_{-\infty}^{\infty} P^\gamma(x-x')f(x')dx' = \int_{-\infty}^{\infty} P_0^\gamma e^{-\frac{\gamma(x-x')^2}{2\sigma^2}}f(x')dx' \qquad (3)$$

Where $f(x)$ is a function describing the profile of the sample, $f(x) = 1$ on the sample and zero otherwise. Equation 3 is the convolution of the Gaussian beam and the photocell profile.

If the maximum of the Gaussian beamspot goes over the critical threshold, $P_0 > P_{th}$, then an electron hole liquid droplet will form near the center of the beamspot, where the power is greater than the critical threshold. Once the droplet forms it can absorb nearby charge carriers. The e-h droplet size should increase linearly as the laser power is increased, thus the size of the droplet $l$ is given by $l = L_0(P_0 - P_{th})$. Where $L_0$ is a free parameter that tunes the rate of expansion of the droplet. We set $L_0$ such that the photocurrent versus power resembles the photocurrent profiles at the center of the sample (i.e. disregarding spatial effects). Since charge carriers inside the droplet recombine, the part of the beamspot that the droplet is under does not contribute to the observed photocurrent, thus the current is:

$$I(x) = \int_{-\infty}^{l} P_0^{\gamma} e^{-\frac{\gamma(x-x')^2}{2\sigma^2}} f(x')dx' + \int_{l}^{\infty} P_0^{\gamma} e^{-\frac{\gamma(x-x')^2}{2\sigma^2}} f(x')dx' \qquad (4)$$

The observed photocurrent is the convolution of the photocurrent due to gas-phase electron-hole pairs and the photocell active area profile. More simply, instead of a Gaussian beamspot convolved with the sample, in the liquid phase, the photoresponse is a "Gaussian donut" convolved with the sample. If we take the equation 4 and plot it as a function of space (main text Figure S3B), we see the characteristic "ring" in the center of the sample emerge above the threshold. The size of the ring is correlated directly with the size of the droplet.

### S4.3 Energetics and Dynamics of the Exciton Gas and Electron-Hole Liquid

Figure S13 shows a schematic of the energy per electron-hole pair in the liquid and gas phases. In the gas phase, excitons in the system are separated by the exciton Bohr radius, $a_B$, and have energy equal to the exciton binding energy $E_{exciton}$. When the charge density increases to the point that the inter-exciton spacing is comparable to the Bohr radius, $a_{xx} \to a_B$, e-h pairs can condense into an electron-hole liquid. The energy difference between the two phases is $\Delta E = E_{liquid} - E_{exciton}$. In the presence of an external electric field, the e-h liquid may be polarized. When the energy produced by the electric field exceeds $\Delta E$, the liquid phase dissociates into bound electron-hole pairs (excitons).

To understand the electron-hole dynamics in the G-MoTe$_2$-G photocell, we consider the number of carriers in three states; $N_f$ the number of free charge carriers, $N_{ex}$ the number of excitons and $N_l$ the number of carriers in the liquid state. Figure S14 shows a schematic that maps out the relevant populations and the transitions between them. Free carriers can become excitons, become part of the electron hole liquid (if it exists), or escape and become observable photocurrent. Excitons can annihilate or form into the liquid state. Electron-hole pairs in the liquid state can evaporate and become free charge carriers or recombine. Radiative recombination in the liquid phase and exciton annihilation in the gas phase result in loss of observable charge carriers and therefore less observed photocurrent. The processes represented as dashed lines are neglected because they are assumed to be rare.

We use this model to write down equations for the rates at which populations shift between the various states from the processes represented as edges on Figure S14. For a pulsed laser, the initial condition is that at time zero $N_0$ carriers are deposited into the sample, i.e. $N_f(t = 0) = N_0$. To consider the two-pulse behavior, we impose the additional condition that at time $\Delta t$ an addition pulse increases $N_f$ by an additional $N_0$ i.e. $N_f(\Delta t) \to N_f(\Delta t) + N_0$. The output of the model is the number of free charge carriers that have escaped and become observable photocurrent, $I_{obs}$. Experimentally, $I_{obs}$ is related to the measured photocurrent signal and $N_0$ is proportional to laser power. This model should give us predictions for photocurrent as a function of power, which are our primary measurements.

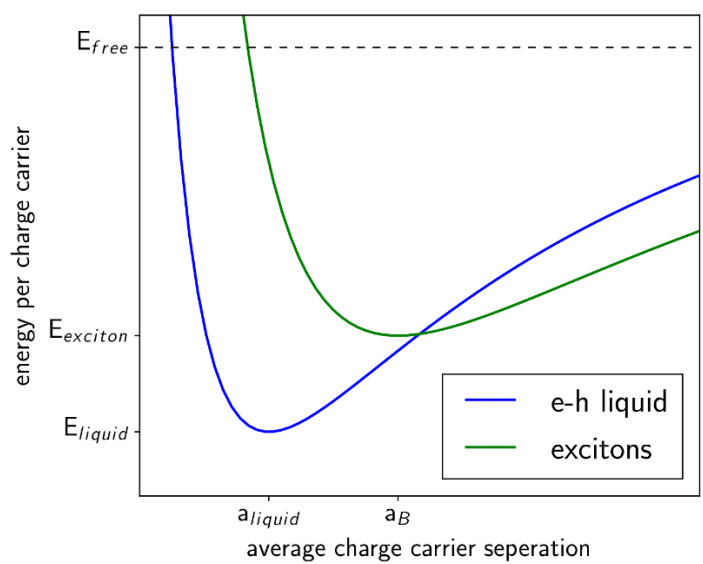

**Fig. S13.**

*Energy diagram of the e-h gas phase and e-h liquid phase.* Schematic energy diagram for photoexcited carriers in MoTe₂ showing the energy curves for the e-h liquid and standard excitons.

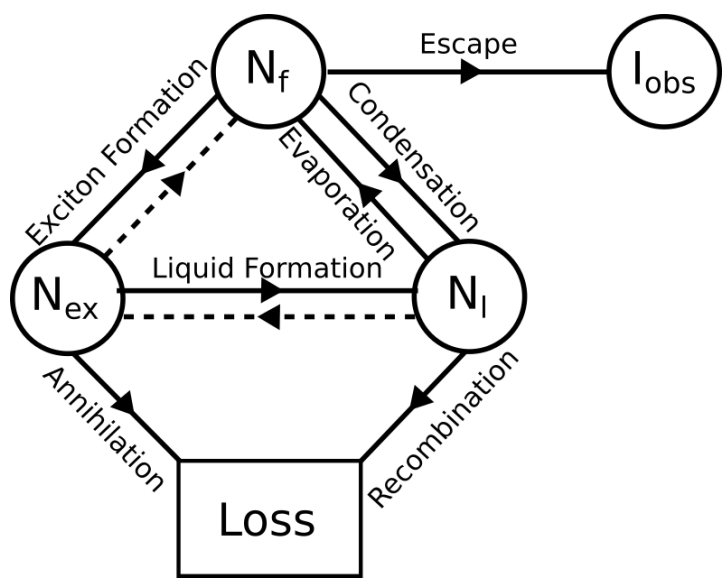

**Fig. S14.**

Charge carrier model. A simple model for calculating the flow of charge carriers between various states.

We first treat the sub-critical photoresponse, for which no condensation has occurred. In the Figure S14 representation, the $N_l$ node and all processes going into or out of it are neglected. In this case the only processes that can happen are free carriers becoming excitons or observable current. The excitons that form can only annihilate and become lost. Experimentally, we cannot measure the observed photocurrent $I_{obs}$ instantaneously, rather photocurrent is measured by the hardware over some integration time which is long compared to the dynamics. Therefore, we define the observed photocurrent signal, as the time integrated current, $PC_{obs} = \int I_{obs} dt$. Thus, the rate equations that give the observed photocurrent are:

$$\frac{dN_f}{dt} = -\alpha N_f^2 - \frac{N_f}{\tau_{es}} \tag{5}$$

$$\frac{dN_{ex}}{dt} = \alpha N_f^2 - \frac{N_{ex}}{\tau_{an}} \tag{6}$$

$$\frac{dPC_{obs}}{dt} = \frac{N_f}{\tau_{esc}} \tag{7}$$

Where exciton formation scales as a two-body process and is parameterized by the constant $\alpha$. The other processes are parameterized by their timescales, $\tau_{es}$ and $\tau_{an}$, the timescales of escape and annihilation respectively. Looking at these equations we see that equation 6 does not have any observable consequences, as carriers that become excitons simply annihilate and are lost.

This case can be solved analytically. First, we integrate equation 5 with the condition that $N_f(0) = N_0$ which gives

$$N(N_0, t) = \frac{N_0 e^{-t/\tau_{es}}}{1 + N_0 \alpha (1 - e^{-t/\tau_{es}})} \tag{8}$$

Then we integrate equation 7 to get $PC_{obs}$

$$PC_{obs} = \int_0^\infty \frac{N_f(N_0, t)}{\tau_{es}} dt = \frac{1}{\alpha \tau_{es}} \ln(1 + N_0 \alpha \tau_{es}) \tag{9}$$

To get the two-pulse behavior as a function of $\Delta t$ the process is similar, but we integrate the first pulse only up to $\Delta t$, then increase the value of $N_f$ by $N_0$ and integrate fully.

$$PC_{obs}(\Delta t) = \int_0^{\Delta t} \frac{N_f(N_0, t)}{\tau_{es}} dt + \int_0^\infty \frac{N_f(N_1, t)}{\tau_{es}} dt, \qquad N_1 = N_0 + N(N_0, \Delta t) \tag{10}$$

Performing this integration gives,

$$PC_{obs}(\Delta t) = \frac{1}{\alpha \tau_{es}} \ln\left[1 + N_0 \alpha \tau_{es}\left(1 - e^{-\Delta t/\tau_{es}}\right)\right] + \frac{1}{\alpha \tau_{es}} \ln(1 + N_1 \alpha \tau_{es}) \tag{11}$$

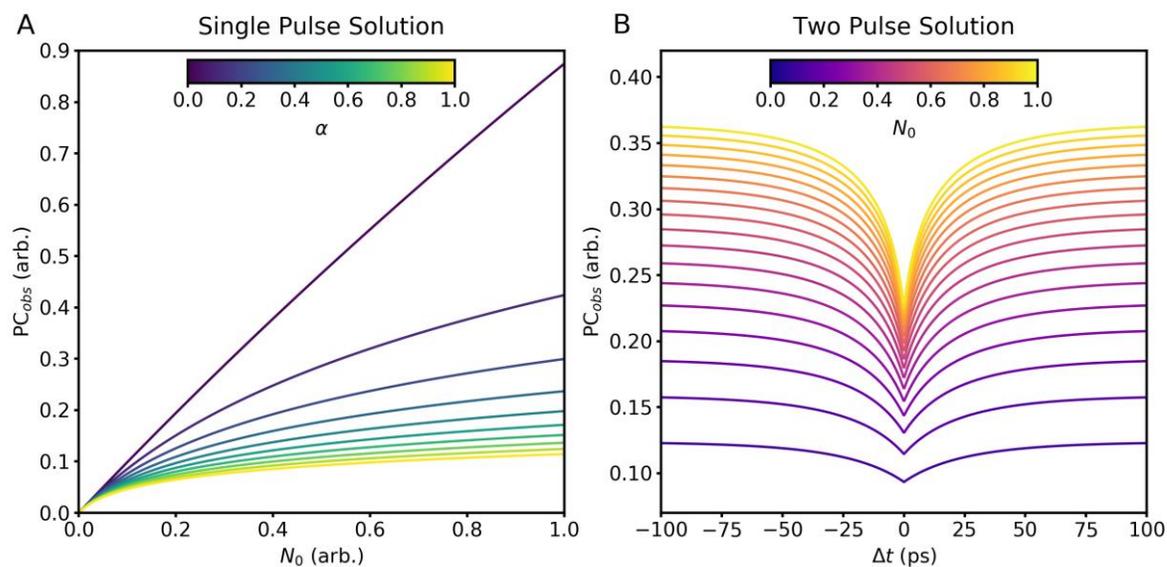

**Fig. S15.**

*Solutions of analytic rate equations.* The observable photocurrent based on equations 9 and 11 (**A** and **B** respectively). In both cases $\tau_{es}$ was arbitrarily chosen to be 30ps and in the two-pulse case $\alpha$ was arbitrarily chosen to be 0.5.

Figure S15 shows the results of these calculations as a function of $N_0$, which is proportional to laser power. In the single pulse case, equation 9, we see that $\alpha$ acts as a tuning parameter for the sub-linearity of the photoresponse, the more exciton formation happens the more sub-linear the photocurrent is. We can also see why our parameterization using $\gamma$ works well, at small $\alpha$ the logarithm can be approximated well by a power law. The two-pulse behavior is calculated from equation 11, and resembles our data at low power. In both cases the photocurrent is monotonically increasing with respect to $N_0$ (power) so we don't expect to see suppression of photocurrent in the absence of an electron hole liquid.

Relaxing this approximation, we consider the case where the density can be high enough for an electron hole liquid to form. For this, we need to consider all the processes shown in Figure S14. To model all the processes, we need to make assumptions about the behavior of the electron hole liquid. Firstly, we need a function that describes the formation of the electron hole liquid from excitons, $\lambda_{form}$. From our data, we know that the liquid forms after a critical point in $N_0$ and then it's volume (and therefore total number) rises linearly. Therefore, we take the following for liquid formation $\lambda_{form}(N_{ex}) = \lambda_0(N_{ex} - N_c)\Theta(N_{ex} - N_c)$. where $N_c$ is the critical density of excitons, $\Theta(N_{ex})$ is a step function and $\lambda_0$ parameterizes the strength of liquid formation.

Our second assumption is how free carriers interact with the electron hole liquid. Initially, the liquid forms by a complex process of nuclear growth. When charge density is injected into the sample microscopic droplet nuclei will form throughout the sample and then merge together into a large-scale droplet. Once that droplet has formed there are two competing processes, evaporation and condensation. Carriers at the surface of the liquid can evaporate off it, becoming free carriers. Or, free carriers can encounter the electron hole liquid and condense onto it. Since our data shows suppression of photocurrent, the coupling between the free carriers and the liquid must decrease the number of free carriers. Since there is no good dynamical model for the complete process of nucleation and condensation we can only model limits of these processes. Thus, we represent the coupling between $N_f$ and $N_l$ as some function $C(S_0, N_f, N_l)$ where $S_0$ parameterizes the strength of the coupling.

In this liquid model, the rate equations take the following form:

$$\frac{dN_f}{dt} = -\alpha N_f^2 - \frac{N_f}{\tau_{es}} - C(S_0, N_f, N_l) \qquad (12)$$

$$\frac{dN_{ex}}{dt} = \alpha N_f^2 - \frac{N_{ex}}{\tau_{an}} - \lambda_0(N_{ex} - N_c)\Theta(N_{ex} - N_c) \qquad (13)$$

$$\frac{dN_l}{dt} = \lambda_0(N_{ex} - N_c)\Theta(N_{ex} - N_c) + C(S_0, N_f, N_l) - \frac{N_l}{\tau_{rec}} \qquad (14)$$

$$\frac{dPC_{obs}}{dt} = \frac{N_f}{\tau_{esc}} \qquad (15)$$

Where $\tau_{rec}$ is the timescale of carrier recombination inside the electron hole liquid. This non-linear system of equations cannot be solved analytically, but we can integrate them numerically. For integration there are several free parameters that need to be chosen, the timescales $\tau_{es}$ and $\tau_{an}$ don't have much effect, beside slightly changing the width of the behavior in the two-pulse simulations, however $\tau_{rec}$ must be much less $\tau_{an}$, because we expect excitons to be relatively stable and that recombination will significantly increase in the highly-correlated electron hole liquid. For the simulations shown in Figures S16 and S17 we chose $\tau_{es} = 30$ ps, $\tau_{an} = 20$ ps and $\tau_{rec} = 1$ ps. The parameters $\alpha$ and $\lambda_0$ were chosen arbitrarily to get a significant number of carriers in the electron hole liquid phase.

We simulate the coupling of the liquid to free carrier for two limits. In the first limit, we assume that the droplet's dynamics are dominated by the formation of the droplet via nucleation and nuclear growth. Nucleation is difficult to model analytically but once droplet nuclei have formed in the sample, they will begin to freely expand and merge together. As the nuclei grow they will gather free carriers from around them into their volume, meaning that their increase in volume depends on the number of surrounding free carriers. This results in coupling that is, to first order, the product of the populations, thus $C(S_0, N_f, N_l) = S_0 N_f N_l$. Simulations performed in this limit are shown in Figure S16 for various values of $S_0$ and $\Delta t$, with parameter values discussed above.

In the other limit, the interactions between free carriers and the liquid occur only after the liquid droplet has formed, and are therefore confined to the surface of the droplet. In this case, there is a cylindrical droplet (for a 2D liquid) surrounded by a volume of free carriers (and excitons). The free carriers can diffuse onto the surface of the droplet. From diffusion onto a cylindrical surface the coupling can be approximated by $C(S_0, N_f, N_l) = S_0 N_f \sqrt{N_l}$. Simulations performed in this limit are shown in Figure S17 for parameter values discussed above. Comparing the two limits, we see that the surface limit shown in Figure S17 involves a much sharper transition as a function of $N_0$. This manifests a sharper suppression of photocurrent in the single pulse simulations and as kinks in the two pulse simulations at moderate values of $N_0$, where the liquid forms at short values of $\Delta t$ but not at long values of $\Delta t$.

Overall, the rate equation model generates results that are qualitatively consistent with our data in either limit. In both limits, as we allow free carriers to interact with the electron hole liquid, the amount of photocurrent is suppressed above some critical point. This is consistent with the suppression that we see in the data for photocurrent, as in main text Figure 2 and supplement Figure S10. In the two pulse results, we see behavior similar to the non-liquid model but with stronger suppression as $\Delta t \to 0$. This is consistent with our data, such as that shown in main text Figure 4B. The truth likely lies somewhere between these two limits, the data has a stronger photocurrent suppression than shown in the nuclear growth limit, but the transition is not as abrupt as the surface limit.

$$C(S_0, N_f, N_l) = S_0 N_f N_l$$

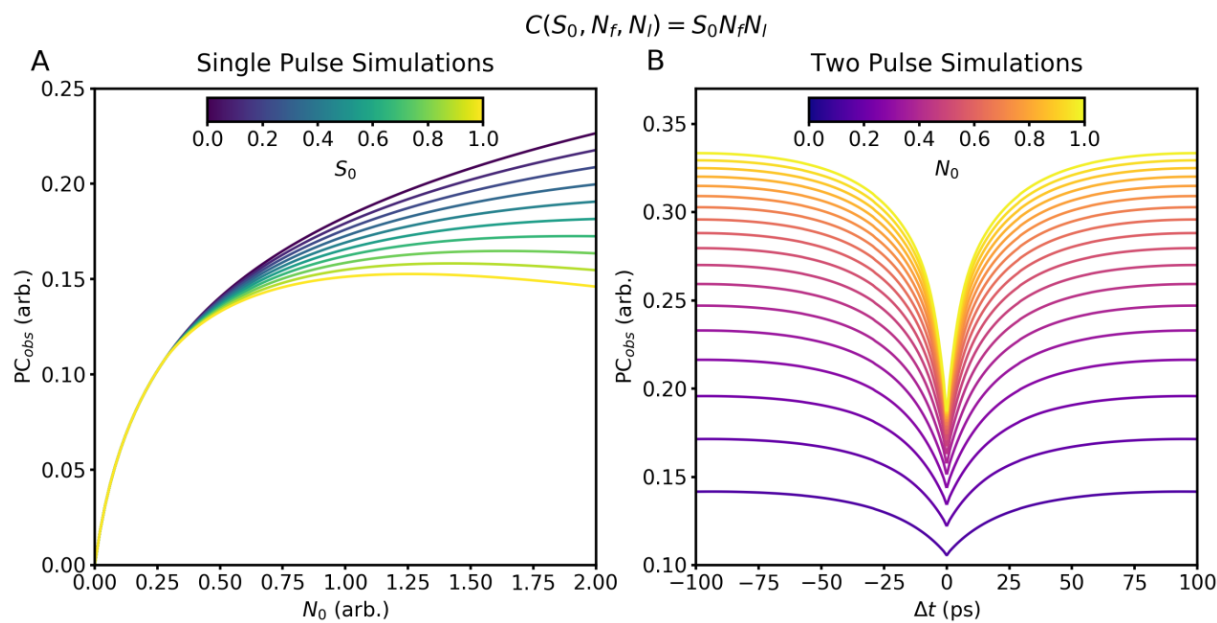

**Fig. S16.**

*Numerical solutions of the full rate equations in the nuclear growth limit.* Simulated photocurrent based on numerical integration of equations 12-15 in the nuclear growth limit where $N_f$ and $N_l$ couple linearly. Parameter values chosen as described in the text.

$$C(S_0, N_f, N_l) = S_0 N_f \sqrt{N_l}$$

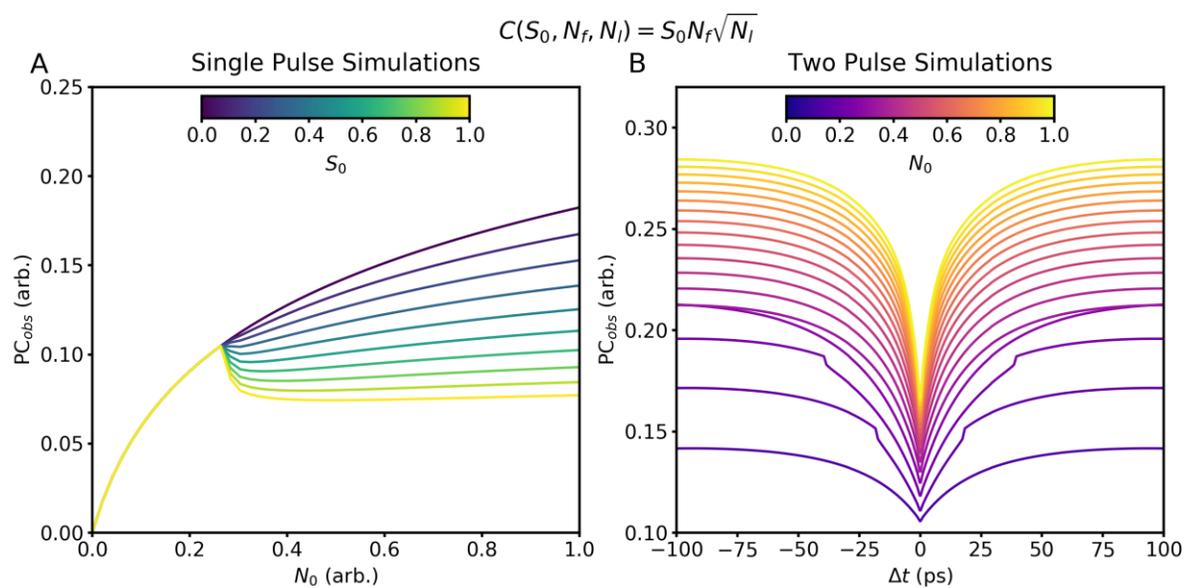

**Fig. S17.**

*Numerical solutions of the full rate equations in the surface limit.* Simulated photocurrent based on numerical integration of equations 12-15 in the limit that the coupling is dominated by condensation at the surface of the droplet, where $N_f$ and $N_l$ couple as $N_f \sqrt{N_l}$. Parameter values chosen as described in the text.